\documentclass[10pt,british,enabledeprecatedfontcommands]{scrartcl}
\usepackage[T1]{fontenc}
\usepackage[utf8]{inputenc}
\usepackage[a4paper]{geometry}
\usepackage{color}
\usepackage{babel}
\usepackage{array}
\usepackage{float}
\usepackage{booktabs}
\usepackage{mathrsfs}
\usepackage{mathtools}
\usepackage{multirow}
\usepackage{amsmath}
\usepackage{amsthm}
\usepackage{amssymb}
\usepackage{graphicx}
\usepackage[authoryear]{natbib}
\usepackage[unicode=true,pdfusetitle,
 bookmarks=true,bookmarksnumbered=false,bookmarksopen=true,bookmarksopenlevel=1,
 breaklinks=false,pdfborder={0 0 1},backref=false,colorlinks=true]
 {hyperref}
\hypersetup{
 citecolor = blue, urlcolor = blue}

\makeatletter

\providecommand{\tabularnewline}{\\}

\@ifundefined{date}{}{\date{}}
\usepackage{mathrsfs}
\usepackage{upgreek}
\usepackage{fixmath}
\usepackage{tikz}
\usepackage{booktabs}
\usepackage{colortbl}
\usepackage{subfigure}
\usepackage[shortcuts]{extdash}

\DeclareMathOperator*{\argmax}{argmax}

\newcommand{\bigperp}{%
  \mathop{\mathpalette\bigp@rp\relax}%
  \displaylimits
}

\newcommand{\bigp@rp}[2]{%
  \vcenter{
    \m@th\hbox{\scalebox{\ifx#1\displaystyle2.1\else1.5\fi}{$#1\perp$}}
  }%
}

\newcommand{\bignparallel}{%
  \mathop{\mathpalette\bignp@rp\relax}%
  \displaylimits
}

\newcommand{\bignp@rp}[2]{%
  \vcenter{
    \m@th\hbox{\scalebox{\ifx#1\displaystyle2.1\else1.5\fi}{$#1\nparallel$}}
  }%
}

\AtBeginDocument{

}

\makeatother

\begin{document}
\global\long\def\NN{\mathbb{N}}%
\global\long\def\P{\mathsf{P}}%
\global\long\def\E{\mathsf{E}}%

\title{CME Iceberg Order Detection and Prediction}
\date{Preprint ver.~2019-08-29}
\author{Dmitry \textsc{Zotikov}\\
Devexperts LLC\\
\texttt{dmitry.zotikov@devexperts.com}
\and
Anton \textsc{Antonov}\\
dxFeed Solutions DE GmbH\\
\texttt{antonov@dxfeed.com}}
\maketitle
\begin{abstract}
We propose a method for detection and prediction of native and synthetic
iceberg orders on Chicago Mercantile Exchange. Native (managed by
the exchange) icebergs are detected using discrepancies between the
resting volume of an order and the actual trade size as indicated
by trade summary messages, as well as by tracking order modifications
that follow trade events. Synthetic (managed by market participants)
icebergs are detected by observing limit orders arriving within a
short time frame after a trade. The obtained icebergs are then used
to train a model based on the Kaplan–Meier estimator, accounting for
orders that were cancelled after a partial execution. The model is
utilized to predict the total size of newly detected icebergs. Out
of sample validation is performed on the full order depth data, performance
metrics and quantitative estimates of hidden volume are presented.
\end{abstract}

\section{Introduction}

On financial exchanges, an \emph{iceberg order} is a limit order where
only a fraction of the total order size (\emph{display quantity})
is shown in the limit order book (LOB) at any one time (\emph{peak}),
with the remainder of volume hidden \citep{Christensen2013}. When
the peak is executed, the next part of the iceberg's hidden volume
(\emph{tranche }or \emph{refill}) gets displayed in the LOB. This
process is repeated until the initial order is fully traded or cancelled.

The hidden volume, although not being directly observed, is de facto
present in the LOB and hence can be traded against. This makes the
detection of hidden liquidity a desirable goal for interested parties,
e.g. traders and market makers.

In this paper we propose a method for detecting and predicting hidden
liquidity on Chicago Mercantile Exchange (CME). The model is fit and
assessed out of sample using historical data. We treat it both as
a classification and a regression model and discuss relevant performance
metrics.

\subsection{Data}

We had access to an almost week-long full order depth (FOD) LOB log
of a September E-Mini S\&P 500 futures contract, existing at that
time under the ticker symbol ESU19, for the period from 2019-06-14,
11:00:00 CDT to 2019-06-21, 16:00:00 CDT. The chosen interval is especially
interesting from a trading activity standpoint: as the front month
contract ESM19 approaches its expiration, a majority of the open interest
gets transferred onto the next one, creating an increased demand for
hidden liquidity vehicles. Each order was described by a sequence
of fields presented in table \ref{tab:lob-log-fields}.

\begin{table}[h]
\begin{centering}
\begin{tabular}{>{\centering}m{0.08\textwidth}>{\raggedright}m{0.12\textwidth}>{\raggedright}m{0.1\textwidth}>{\raggedright}m{0.1\textwidth}>{\raggedright}m{0.18\textwidth}>{\raggedright}m{0.12\textwidth}>{\raggedright}m{0.12\textwidth}}
\toprule 
 & Time & Order ID & Side & Action & Price & Volume\tabularnewline
\midrule
Possible values & Millisecond resolution & 12-digit identifier & ``B'' (buy), ``S'' (sell) & ``Limit'' (new), ``Modify'' (update), ``Delete'' & Non-negative real & Non-negative integer\tabularnewline
\bottomrule
\end{tabular}
\par\end{centering}
\caption{\label{tab:lob-log-fields}Order log fields for the ESU19 data.}
\end{table}

For more information about CME Market-by-Order book management, see
\citep{CmeMboBookManagement}.

In addition, trade summary messages \citep{CmeTradeSummary} were
present in the data. Each trade event against a resting order corresponded
to a trade record in the log with the aforementioned fields, ``Action''
set to ``Trade'' and an extra field for the passive order ID.

This way, our algorithm works in the ``offline'' mode by reading
a pre-recorded LOB log. Since it is not forward-looking, it can be
easily modified to work with real-time streaming data.

\subsection{CME Iceberg Orders}

CME supports two types of iceberg orders: \emph{native} and \emph{synthetic}
\citep{CmeMbo}.
\begin{description}
\item [{Native}] icebergs are managed by the exchange itself. All new tranches
are submitted as modifications of the initial order; this means that
the original order ID is preserved throughout the whole lifetime of
the iceberg. Additionally, trades against these orders may sometimes
be larger in volume than the current resting size, as indicated by
trade summary messages.
\item [{Synthetic}] icebergs are submitted by independent software vendors
(ISV), whose infrastructure is physically separated from the exchange.
ISV's split the initial iceberg order, submit new tranches and track
their execution. These tranches are indistinguishable from usual limit
orders submitted by other participants.
\end{description}
Detecting native icebergs is conceptually easy since a) the order
ID does not change until the iceberg is fully executed or cancelled;
and b) trade summary messages include actual trade volumes, which
may be larger than the resting display quantity. Thus an unambiguous
and accurate detection is possible. Synthetic icebergs, on the other
hand, being identical to non-iceberg orders in how they are processed
by the exchange, can only be detected heuristically and relying on
a set of assumptions, which are introduced further.

\section{Existing Literature Overview\label{sec:Existing-Literature-Overview}}

For a good literature overview see \citep[p.\,7]{Christensen2013}.
In particular, some authors had access to order logs in which iceberg
orders were explicitly identified, so that an unambiguous reconstruction
of both displayed and hidden LOB's was possible, e.g. by \citep{Frey2009}
for XETRA exchange. There are a few articles that provide hidden liquidity
estimates — see \citep[p.\,5]{Hautsch2010} and \citep[p.\,7]{Christensen2013}
for such lists. These estimates differ between authors, exchanges
and instrument types, ranging from 2\% \citep{FlemingMizrach2008}
to 52\% \citep{Moro2009}. That being said, most of the papers were
published almost a decade ago, so one may argue that contemporary
markets have different hidden volume properties.

To our knowledge, there are virtually no articles that have the premise
of simultaneous iceberg detection and prediction in the same setting
as ours, with the exception of \citep{Christensen2013}, who propose
a solution to the very problem which the authors of the present paper
are concerned with. Their approach consists of the following three
phases:
\begin{enumerate}
\item Series of tranches are identified in the data as belonging to larger
iceberg orders (the ``\emph{detection}'' step).
\item Using the detected icebergs, a statistical model is fit that captures
the correspondence between the peak size and the total iceberg size
(the ``\emph{learning}'' step).
\item The detection step is repeated, but a new iceberg order is detected,
a prediction of the total size of the iceberg is made using the model
obtained at the previous step (the ``\emph{prediction}'' step).
\end{enumerate}
We adapt this detection–learning–prediction scheme for our work, albeit
with the following notable differences:
\begin{itemize}
\item The authors did not have the access to the FOD MBO data at the time
of writing. In particular, the order ID for each action or trade was
not available, yet that drastically changes the logic of the detection
step.
\item No distinction between synthetic and native icebergs is made. Namely,
it is assumed that trades can sometimes be larger than the size of
the resting order being traded, what is specific for native icebergs;
however, iceberg tranches arrive as new limit orders, and that is
an attribute of synthetic icebergs.
\item During the learning phase, a bivariate Gaussian kernel density estimate
of peak and total size is built, which is then optimised for the global
maximum given a peak size. For the purpose of prediction, where only
one value of the total size corresponding to the maximum probability
given a peak size is necessary, this complication is questionable
as a simpler model is sufficient\footnote{It should be noted that the authors consider discrete kernel estimation,
but opt to use the Gaussian kernel ``on the basis of simplicity''.}. Kernel density estimate may be desired if the algorithm operates
on instruments with a relatively low daily trading volume, and this
is not the case with our data. In addition, by omitting this step
we don't have to resort to numerical methods when optimizing for conditional
maxima.
\item All incomplete icebergs — i.e. those, that were cancelled before being
fully executed — are not included into the learning phase. However,
our calculations show that more than half of all synthetic icebergs
are cancelled, thus it is highly desirable to include the information
about incomplete executions into the model.
\end{itemize}

\section{Detection}

\subsection{Native Icebergs}

Native iceberg orders enter the book as limit orders which may or
may not be traded upon arrival. After the initial limit order volume
is fully traded, the next part of the iceberg order appears in the
book. Crucially, when the iceberg has its displayed quantity refreshed
(by means of an update action), the refreshed order will have the
same order ID as the original order. Moreover, any trades involving
the iceberg order will indicate the total volume of trade, including
the hidden part of the iceberg. Using these two properties it is then
fairly easy to detect a sequence of new–trade–update–delete actions
that forms an iceberg. In particular, we might be interested in update
actions that correspond to new iceberg tranches, as well as in determining
the peak size and in calculating the total iceberg size.

We would like to illustrate the process with an example. Consider
the data presented in table \ref{tab:Sample-native-iceberg-log}.

\begin{table}[h]
\renewcommand{\arraystretch}{1}

\begin{table}[H]
\centering
\begin{tabular}{lrllrrr}
\toprule
Time & Order ID & Side & Action & Price & Volume & Affected\\
\midrule
14:05:33.416 & 645764830354 & S & Trade & 2931.75 & 2 & 645764830338\\
14:05:33.416 & 645764830354 & S & Trade & 2931.75 & 10 & 645764830339\\
14:05:33.416 & 645764830354 & S & Limit & 2931.75 & 6 & -\\
14:05:33.416 & 645764830360 & B & Trade & 2931.75 & 8 & 645764830354\\
\addlinespace
14:05:33.416 & 645764830354 & S & Modify & 2931.75 & 7 & -\\
14:05:33.416 & 645764830361 & B & Trade & 2931.75 & 3 & 645764830354\\
14:05:33.416 & 645764830354 & S & Modify & 2931.75 & 4 & -\\
14:05:33.416 & 645764830362 & B & Trade & 2931.75 & 2 & 645764830354\\
14:05:33.416 & 645764830354 & S & Modify & 2931.75 & 2 & -\\
14:05:33.416 & 645764830363 & B & Trade & 2931.75 & 1 & 645764830354\\
14:05:33.416 & 645764830354 & S & Modify & 2931.75 & 1 & -\\
14:05:33.416 & 645764830365 & B & Trade & 2931.75 & 1 & 645764830354\\
\addlinespace
14:05:33.416 & 645764830354 & S & Modify & 2931.75 & 9 & -\\
14:05:33.416 & 645764830366 & B & Trade & 2931.75 & 1 & 645764830354\\
14:05:33.416 & 645764830354 & S & Modify & 2931.75 & 8 & -\\
14:05:33.416 & 645764825841 & B & Trade & 2931.75 & 1 & 645764830354\\
14:05:33.416 & 645764830354 & S & Modify & 2931.75 & 7 & -\\
14:05:33.417 & 645764830382 & B & Trade & 2931.75 & 9 & 645764830354\\
\addlinespace
14:05:33.417 & 645764830354 & S & Modify & 2931.75 & 5 & -\\
14:05:33.417 & 645764830390 & B & Trade & 2931.75 & 5 & 645764830354\\
14:05:33.417 & 645764830354 & S & Delete & 2931.75 & 5 & -\\
\bottomrule
\end{tabular}
\end{table}

\renewcommand{\arraystretch}{1}

\caption{\label{tab:Sample-native-iceberg-log}Sample native iceberg order
log data. Grouped are the orders related to the same tranche.}
\end{table}

\begin{enumerate}
\item Order \#645764830354 enters the book and immediately gets traded at
2931.75 for the total of 12 units of volume. The remainder — 6 units
— is placed at the same level. At that point we do not know whether
the order has any hidden liquidity or not. Moreover, assuming that
it does, the peak size cannot be precisely determined; but since $12+6=18$,
it is one of the divisors of 18 greater or equal than 6, i.e.~6,~9~or~18.
\item The next trade has volume 8 which is larger than the resting volume
of 6. This is sufficient to mark order \#645764830354 as an active
iceberg.
\item The next tranche volume is 7. Note that $8-6=2$ units of volume were
traded against a tranche that had not entered the book. This means
that the peak size can be determined precisely as $7+(8-6)=9$. The
trade could have been large enough to consume several hidden tranches.
\item The next several trades are smaller in volume than the resting order.
The trade initiated by order \#645764830365 is equal to the resting
volume of 1. Consequently, the next modify action is seen to refresh
the visible volume by the peak size of 9 (which agrees with the previous
calculations).
\item Finally, the last update action has volume 5, which, accounting for
the hidden trade of $9-7=2$ results in peak size of 7. The trade
for 5 units completes the sequence as no more refresh messages is
seen and the order is deleted from the book.
\end{enumerate}
Overall, the iceberg has a total volume of 43, 4 tranches with peak
sizes 9, 9, 9 and 7, correspondingly, and the display quantity equal
to 9.

\begin{figure}[H]
\begin{centering}
\includegraphics[width=1\textwidth]{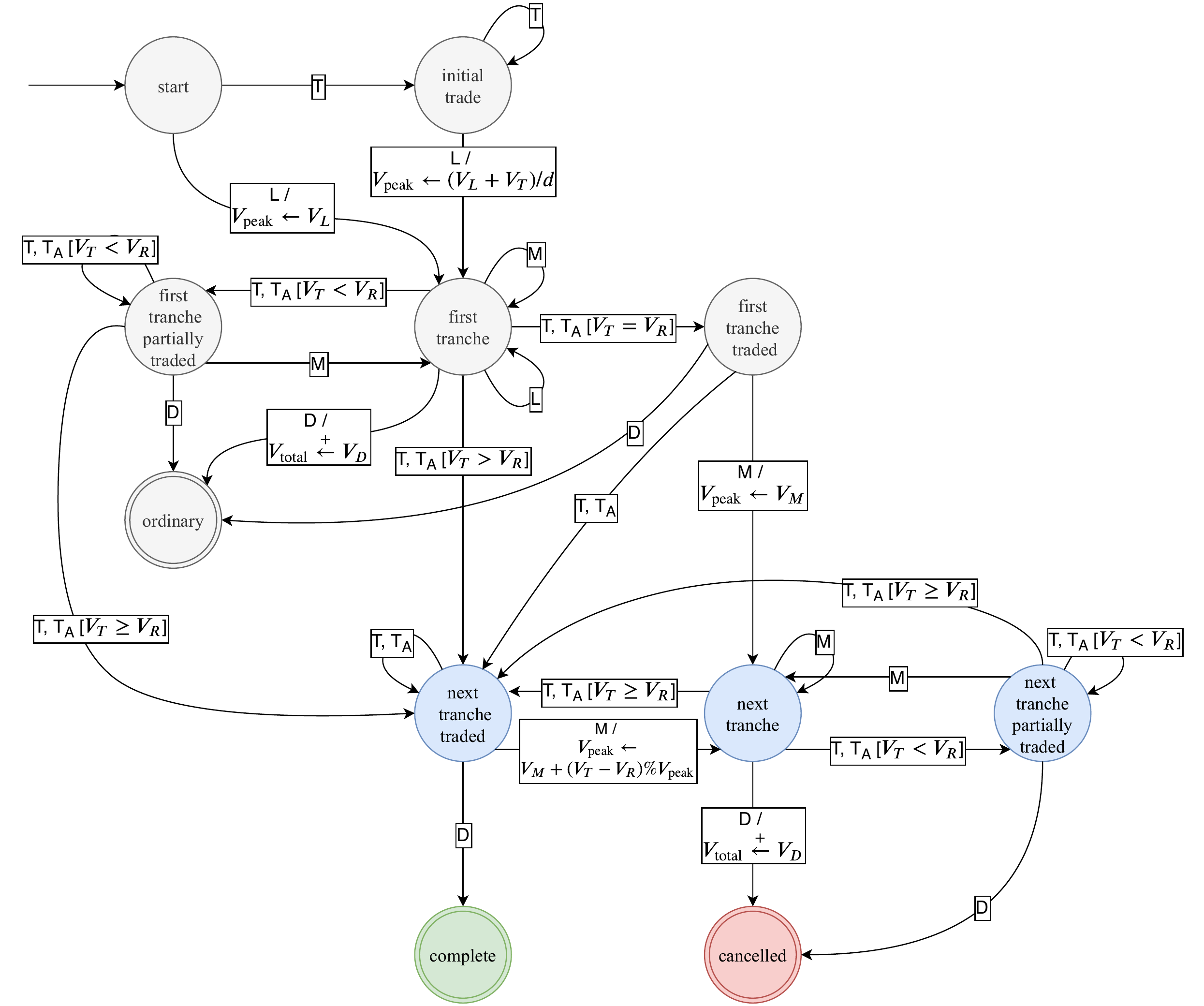}
\par\end{centering}
\caption{\label{fig:iceberg-fsa}The grammar of native icebergs. The nodes
correspond to states of the finite state machine, the edges~—~to
order actions: new (\texttt{L}), update (\texttt{M}), trade (\texttt{T}),
``affected'' trade (\texttt{T}\protect\textsubscript{\texttt{A}}),
delete (\texttt{D}). Trades~\texttt{T} are initiated by the iceberg
order, while ``affected'' trades~\texttt{T}\protect\textsubscript{\texttt{A}}~—~by
other incoming orders. An optional condition is specified in square
brackets; a side effect is specified after the slash (\texttt{/}).
Different volumes~$V_{\cdot}$ refer to the iceberg's peak volume
$V_{\text{peak}}$, total volume $V_{\text{total}}$, current resting
volume $V_{R}$ and the order's trade ($V_{T}$), delete ($V_{D}$),
modify ($V_{M}$) volumes. Node colours represent the status of the
action sequence being tracked: grey for non-iceberg (``ordinary''),
blue for active (all statuses starting with ``next tranche...''),
green for complete and red for cancelled.}
\end{figure}

The process of parsing the action stream can be conveniently formalised
as a finite state machine, see fig. \ref{fig:iceberg-fsa}.

To recap the detection phase, an iceberg enters the book as a new
limit order, possibly following a sequence of trades. It is then traded,
and usually — but not always — each trade corresponds to one trade
summary message, in which case it is followed by an update action,
specifying the currently resting order volume. If more than one trade
messages are seen before the next update action, then this should
be accounted for. Moreover, all price adjustments which move the order
to the top of the book are not disseminated by the exchange, meaning
that even after the placement the order can again act as an aggressive
order and initiate a trade. If at this point the order is deleted
from the book or traded so that the trade volume is never greater
than the resting volume, it is marked as ``ordinary'' and removed
from consideration. On the other hand, once a trade larger than the
resting volume is detected, or the order is fully traded but then
modified to have non-zero volume again, then the order is marked as
an iceberg. The trade–modify cycle then continues until the order
is completely executed or cancelled, resulting in its deletion from
the book.

In addition to tracking the transitions through the state space, we
are interested in calculating the following quantities:
\begin{itemize}
\item \emph{the total volume} $V_{\text{total}}$ is conveniently computed
as the sum of all traded volume $V_{T}$ (which may exceed the sum
of limit and/or update volumes), plus any volume $V_{D}$ that is
explicitly deleted;
\item \emph{the currently resting volume} $V_{R}$ is simply the last modify
action volume $V_{M}$;
\item \emph{the peak size} $V_{\text{peak}}$ is determined iteratively:
\begin{itemize}
\item If the iceberg order enters the book directly as a limit order, then
the limit order volume is indeed the peak size.
\item If a series of trades precedes the limit order placement, then $V_{L}=V_{\text{peak}}-V_{T}\bmod V_{\text{peak}}$.
Therefore, $V_{T}-kV_{\text{peak}}=V_{\text{peak}}-V_{L}$ for some
$k\in\NN_{0}$, from which we get 
\[
V_{\text{peak}}=\frac{V_{T}+V_{L}}{k+1}=\frac{V_{T}+V_{L}}{d},\quad d=1,\ldots,V_{T}+V_{L},\ V_{\mathrm{peak}}\in\{n\in\mathbb{N}:n\geq V_{L}\}.
\]
 If more than one admissible $V_{\text{peak}}$ values are found,
then the following heuristics apply.
\begin{itemize}
\item If the first tranche is traded for exactly the resting volume, the
following update message unambiguously identifies the peak size.
\item If the trade volume is greater than the resting volume, then 
\[
V_{\text{peak}}=V_{M}+(V_{T}-V_{R})\bmod V_{\text{peak}}^{*},
\]
where $V_{\text{peak}}^{*}$ is one of previously computed values.
Only the values that satisfy this equation are~kept.
\end{itemize}
\end{itemize}
\end{itemize}

\subsection{Synthetic Icebergs}

Unlike native icebergs, synthetic icebergs are managed by ISVs. After
a tranche is fully executed, the ISV is expected to refill the display
quantity with a new limit order. Since the ISV infrastructure is located
outside the exchange, a short non-constant delay (dubbed \emph{$dt$})
is expected between the corresponding ``delete'' and ``new'' actions.
This idea is also exploited in \citep{Christensen2013}, although
their explanation was different. Additionally, we assume that new
tranches arrive to the same price level as the initial tranche, and
their volumes are expected to be equal to the initial tranche volume
as well, which is taken to be the iceberg display quantity. It should
be reiterated that we only detect orders that have constant peak size.
Under the current model it is impossible to detect synthetic icebergs
with varying peak sizes or price levels. This in particular implies
that if the iceberg is not a multiple of the display quantity, the
last tranche will be smaller than all the previous tranches in volume,
hence its detection using the current approach does not seem to be
possible.

If a tranche is executed, but no refill orders follow within $dt$,
the iceberg is considered \emph{complete}. If a tranche is placed
and later cancelled, the whole iceberg is considered cancelled (\emph{incomplete}).

\begin{figure}[H]
\begin{centering}
\includegraphics[width=0.5\textwidth]{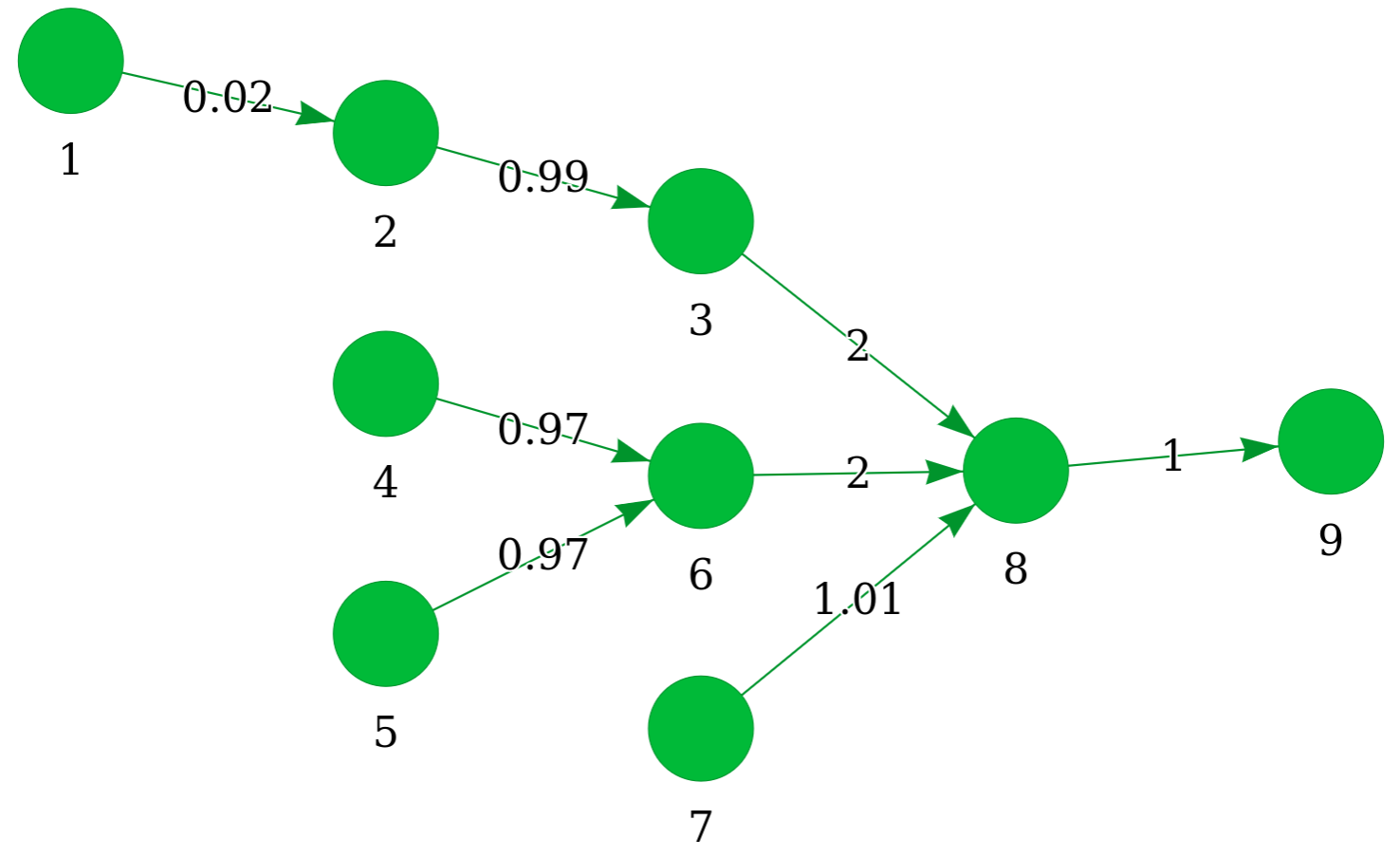}
\par\end{centering}
\caption{\label{fig:test-tranche-tree}An iceberg tranche tree corresponding
to table \ref{tab:Artificial-data-synthetic}. Node labels are order
IDs. Edge labels are time in seconds between subsequent tranches —
note that these are different from $dt$ as a tranche can remain indefinitely
long in the book after its placement. The iceberg consists of either
3, 4 (two chains) or 5 tranches.}
\end{figure}

One complication is that if the activity in the LOB is high (as in
the case with E-Mini S\&P 500 contracts), more than one order of the
target volume may arrive on the same price level within $dt$. Our
very strong assumption is that the next tranche arrives faster than
any other new limit order, so for each tranche there is only one child.
A more sophisticated model would account for all possible children
tranches and somehow average the volume later on. Another complication
is that when several limit orders of the same price and size get executed
and deleted from the book simultaneously, the next tranche can be
``linked'' to any of those. Repeated over several trades, this produces
a \emph{tree} of possible tranches. Every path from all leaves to
the root (a \emph{chain}) is a possible iceberg. See table \ref{tab:Artificial-data-synthetic}
and the resulting graph in fig.~\ref{fig:test-tranche-tree} for
an illustration.

\begin{table}[h]
\begin{table}[H]
\centering
\begin{tabular}{lrllrrr}
\toprule
Time & Order ID & Side & Action & Price & Volume & Affected\\
\midrule
18:22:12.00 & 1 & S & Limit & 1000 & 2 & -\\
\addlinespace
18:22:12.01 & 101 & B & Trade & 1000 & 2 & 1\\
18:22:12.01 & 1 & S & Delete & 1000 & 2 & -\\
18:22:12.02 & 2 & S & Limit & 1000 & 2 & -\\
\addlinespace
18:22:12.04 & 4 & S & Limit & 1000 & 2 & -\\
\addlinespace
18:22:12.04 & 5 & S & Limit & 1000 & 2 & -\\
\addlinespace
18:22:13.00 & 102 & B & Trade & 1000 & 2 & 2\\
18:22:13.00 & 2 & S & Delete & 1000 & 2 & -\\
18:22:13.01 & 3 & S & Limit & 1000 & 2 & -\\
\addlinespace
18:22:13.00 & 103 & B & Trade & 1000 & 2 & 4\\
18:22:13.00 & 104 & B & Trade & 1000 & 2 & 5\\
18:22:13.00 & 4 & S & Delete & 1000 & 2 & -\\
18:22:13.00 & 5 & S & Delete & 1000 & 2 & -\\
18:22:13.01 & 6 & S & Limit & 1000 & 2 & -\\
\addlinespace
18:22:14.00 & 7 & S & Limit & 1000 & 2 & -\\
\addlinespace
18:22:15.00 & 105 & B & Trade & 1000 & 2 & 3\\
18:22:15.00 & 106 & B & Trade & 1000 & 2 & 6\\
18:22:15.00 & 107 & B & Trade & 1000 & 2 & 7\\
18:22:15.00 & 3 & S & Delete & 1000 & 2 & -\\
18:22:15.00 & 6 & S & Delete & 1000 & 2 & -\\
18:22:15.00 & 7 & S & Delete & 1000 & 2 & -\\
18:22:15.01 & 8 & S & Limit & 1000 & 2 & -\\
\addlinespace
18:22:16.00 & 108 & B & Trade & 1000 & 2 & 8\\
18:22:16.00 & 8 & S & Delete & 1000 & 2 & -\\
18:22:16.01 & 9 & S & Limit & 1000 & 2 & -\\
\addlinespace
18:22:16.50 & 109 & B & Trade & 1000 & 2 & 9\\
18:22:16.50 & 9 & S & Delete & 1000 & 2 & -\\
\bottomrule
\end{tabular}
\end{table}

\caption{\label{tab:Artificial-data-synthetic}Artificial data to demonstrate
synthetic iceberg detection. For this example, $dt$ was set to 0.3
seconds. Limit order \#1 gets traded and removed from the book. The
following limit order \#2 arriving within a third of a second becomes
the next tranche in the iceberg chain. Note that orders \#4 and \#5
do not arrive within $dt$ and, since there were no more trades, start
two new chains. After they get traded simultaneously, order \#6 arrives
within $dt$, thus becoming the next tranche. The process continues
until all orders are removed from the book.}
\end{table}

We are also interested in computing the following related quantities:
\begin{itemize}
\item \emph{the peak size} $V_{\text{peak}}$ is trivially set to be equal
to the volume of the initial tranche;
\item \emph{the total volume} $V_{\text{total}}$ could have been calculated
as the sum of the tranche volumes if there was only one tranche chain
per iceberg. In general, however, there are more and the total volume
has to be aggregated in some way. We propose the following options:
\begin{itemize}
\item the average total volume of all chains $\mathcal{V}^{\text{all}}$;
\item the average total volume of chains of unique length $\mathcal{V}^{\text{unique}}$;
\item the total volume of the longest chain $\mathcal{V}^{\text{longest}}$.
\end{itemize}
\end{itemize}

\section{\label{sec:Learning}Learning}

\subsection{Kaplan–Meier estimation}

Having detected sufficiently many iceberg orders, we would like to
build a model that yields a prediction of the total iceberg size.
Although it is clearly not the most advanced in terms of predictive
power, we elaborate on the model proposed in \citep{Christensen2013}.
Namely, for each of $P$ unique detected peak sizes, a distribution
of possible total sizes is estimated; then, given the peak size $V_{\text{peak}}=p$
of a previously unseen iceberg, ``the best'' total volume (in terms
of conditional mean, median or mode) is returned as a prediction.

More precisely, from now on let $V_{p}$ denote a random variable
representing the total volume of an iceberg with peak size $p$. Then
for each value of $p$ we are interested in estimating the distribution
of $V_{p}$. While a trivial empirical distribution might suffice,
our experiments show that a significant amount of synthetic icebergs
are cancelled before being completely executed (see fig.~\ref{fig:Iceberg-completion-dist}).
Hence for some icebergs only a lower bound on their total volume is
known: for the $i$-th iceberg, $v_{i}\geq c_{i},\ c_{i}\in\NN$.
From the point of view of survival analysis, these are \emph{censored}
observations. Usually survival analysis deals with the so-called ``time
to event data'': the primary interest is the time until the onset
of an event for each member of the analysed group. If only upper (or
lower, or both) bounds on time, but not exact event times, are known,
the observations are considered censored. Instead of discarding these,
it is possible to construct estimators that incorporate the uncertainty
associated with censoring. In our case, accumulated iceberg volumes
play the role of time to event durations, so the task is to estimate
the distribution of $V_{p}$ for each $p$ using random right-censored
data.

The proportion of cancelled native icebergs is much smaller, and,
in fact, could be disregarded for the purpose of distribution estimation.
 Nevertheless, we would like to utilise the same approach to simplify
the analysis and to make the direct comparison between native and
synthetic iceberg estimates possible.

The standard approach for a non-parametric distribution estimation
of censored data is to use the \emph{Kaplan–Meier estimate} \citep{Kaplan58}.
Let $F_{p}(v)$ be the cumulative distribution function of $V_{p}$,
then $S_{p}(v)=1-F_{p}(v)$ is its \emph{survival function}. Also
define (for the given $p$)
\begin{labeling}{00.00.0000}
\item [{$u_{1},\ldots,u_{K}$}] unique volumes of all detected icebergs
sorted in ascending order,
\item [{$d_{j}$}] the number of complete icebergs of volume $u_{j}$,
where $j=1,\ldots,K$,
\item [{$n_{j}$}] the total number of both complete and incomplete icebergs
of volumes $u_{j},\ldots,u_{K}$.
\end{labeling}
Then from the general theory of survival analysis it is known \citep{KalbfleischPrentice200209}
that the maximum likelihood estimate of $S_{p}$ is 
\begin{equation}
\hat{S}_{p}(v)=\prod_{j:u_{j}\geq v}\left(1-\frac{d_{j}}{n_{j}}\right).\label{eq:km-estimate}
\end{equation}

\subsection{Weighted Kaplan–Meier estimation for synthetic icebergs}

For synthetic icebergs estimation (\ref{eq:km-estimate}) cannot be
performed directly, because, given a tranche tree, there is no unambiguous
way to calculate the total iceberg size. Instead, we propose a weighting
scheme that assigns weights to each chain within a tranche tree. Given
the $i$-th tranche tree with $h_{i}$ chains of unique length, the
weights are 
\[
w_{i,\ell}=1/h_{i},\quad\ell\in1:h_{i}.
\]
A weight can be also interpreted as a probability of the total iceberg
volume being equal to the accumulated tranche chain volume. Speaking
in terms of Bayesian inference, we assign uniform probabilities because
there is no prior knowledge that would affect our preference for a
particular chain. Then for the purpose of calculating $\hat{S}_{p}(v)$,
instead of $d_{j},n_{j}$, their weighted counterparts $\tilde{d}_{j}$,
$\tilde{n}_{j}$ are computed. Formally, given the definition $\left\{ u_{j}\right\} _{j=1,\ldots,K}$
above, let $\{v_{i,\ell}\}_{\ell=1,\ldots,h_{i}}$ denote a set of
unique volumes of the $\ell$-th chain of the $i$-th tranche tree.
Then
\[
\tilde{d}_{j}=\sum_{i\in C}\sum_{\ell\in H_{i}}w_{i,\ell},\quad H_{i}=\left\{ \ell:v_{i,\ell}=u_{j}\right\} ,
\]
where $C$ is a set of indices of all complete icebergs and $H_{i}$
is a set of tranche chain indices of the $i$-th iceberg, having total
volumes equal to $u_{j}$. $\tilde{n}_{j}$ are computed similarly.
Of course, when each tranche tree consists of only one chain, all
weights are equal to 1 and we have $\tilde{d}_{j}=d_{j}$ and $\tilde{n}_{j}=n_{j}$.
Since $V_{p}$ only takes discrete values for all $p$, we finally
obtain the weighted estimate
\[
\hat{S}_{p}(u_{j})=\prod_{k=1}^{j}\left(1-\frac{\tilde{d}_{k}}{\tilde{n}_{k}}\right).
\]

From $\hat{S}_{p}$ an estimate of the probability mass function $f_{p}(u_{j})=\P(V_{p}=u_{j})$
can be obtained in a trivial way. One notable problem with this estimate
is that if $d_{K}=0$, then $S_{p}(u_{K})\neq0$ and the probabilities
do not sum up to 1. This is fixed trivially by normalising the probabilities.

\section{Prediction}

The prediction step starts from detecting first several tranches of
an iceberg: for native icebergs, this might be any moment when the
iceberg becomes ``active'', for synthetic icebergs this number is
an algorithm parameter with a default value of 3. If the peak size
$p$ is precisely detected, a prediction of the total volume might
be done.

\subsection{Native Icebergs}

For native icebergs we make predictions in terms of the conditional
mean and median. In addition, a different prediction is made based
on 3 highest conditional probabilities. For a fixed iceberg, let $v_{r}$
denote the currently accumulated volume up to, but not including,
tranche number $r$ and let $\mathscr{V}_{p}=\left\{ u_{j}:u_{j}>v_{r}\right\} _{j=1,\ldots,K_{p}}$
be the constrained optimization space. Then define
\begin{itemize}
\item \emph{mean} prediction based on 
\[
\E(V_{p}\mid V_{p}>v_{r})=\frac{1}{\P(V_{p}>v_{r})}\sum_{u\in\mathscr{V}_{p}}u\,\P(V_{p}=u)=\left(\sum_{u\in\mathscr{V}_{p}}f_{p}(u)\right)^{-1}\sum_{u\in\mathscr{V}_{p}}uf_{p}(u)
\]
and defined as
\[
\hat{v}^{\text{mean}}=\left(\sum_{u\in\mathscr{V}_{p}}\hat{f}_{p}(u)\right)^{-1}\sum_{u\in\mathscr{V}_{p}}u\hat{f}_{p}(u),
\]
rounded to the nearest integer;
\item \emph{median} prediction as 
\[
\hat{v}^{\text{median}}=\max\left\{ u_{J}:\sum_{j=1}^{J}\hat{f}_{p}(u_{j})\leq0.5,\quad u_{j}\in\mathscr{V}_{p}\;\forall j=1,\ldots,\left|\mathscr{V}_{p}\right|\right\} ;
\]
\item $k$ \emph{best} \emph{mode} predictions as the $k$-th order statistic
\[
\hat{v}^{\text{mode}(k)}=u_{(k)},
\]
where the order of $u_{(1)},\ldots,u_{(\left|\mathscr{V}_{p}\right|)}$
is given by $\hat{f}_{p}(u_{(1)})\geq\dots\geq\hat{f}_{p}(u_{(\left|\mathscr{V}_{p}\right|)})$.
Tied volumes are taken in ascending order.
\end{itemize}

\subsection{Synthetic Icebergs}

Once again, consider a fixed iceberg and an associated tranche tree.
Let $v_{\ell,r}^{\prime}$ denote the currently accumulated volume
for the $\ell$-th chain up to and including tranche number $r$.
Then our prediction is 
\[
\hat{v}_{\ell}^{\text{mode}}=\argmax_{u\in\mathscr{V}_{p}^{\prime}}\hat{f}_{p}(u),\quad\mathscr{V}_{p}^{\prime}=\left\{ u_{j}:u_{j}\geq v_{\ell,r}^{\prime}\right\} _{j=1,\ldots,K_{p}},
\]
where $\mathscr{V}_{p}^{\prime}$ is the constrained optimization
space and $K_{p}$ is the number of unique total icebergs sizes with
peak~$p$. For the sake of brevity we do not report other possible
estimates, as they do not differ much. The predicted total volume
$\hat{v}_{\ell}^{\text{mode}}$ is aggregated across the chains of
the iceberg in question as 
\begin{itemize}
\item the average total volume of all chains $\hat{\mathcal{V}}^{\text{all}}$;
\item the average total volume of chains of unique length $\hat{\mathcal{V}}^{\text{unique}}$;
\item the total volume of the longest chain $\hat{\mathcal{V}}^{\text{longest}}$.
\end{itemize}

\section{Evaluation}

Given an estimate of $\hat{f}_{p}(v)$ and previously unseen data,
the model can be evaluated both as a binary classifier and as a regression.
In the discussion below, we assume that the prediction algorithm was
run, producing a set of complete icebergs.
\begin{itemize}
\item For \emph{classification}, our null hypothesis is ``there is no hidden
liquidity''. In the context of synthetic icebergs, it means that
the iceberg is complete and no more tranches will follow. In the context
of native icebergs, it means the last seen tranche can only be traded
for the volume not exceeding its currently visible volume, and that
no more tranches will follow. Since the full information on a particular
iceberg execution is available after we run the prediction algorithm
(each iceberg is eventually complete), the true total volume is known\footnote{For synthetic icebergs — only in terms of our model.}
and hence the classification results can be summarised in a confusion
matrix, from which we compute the standard classification metrics:
accuracy, precision, recall and F1 score.
\item \emph{Regression} performance metrics show the degree to which the
prediction is different from the true total volume.
\end{itemize}
The details of evaluation are slightly different for native and synthetic
icebergs, and are given below. We hope that the level of details is
sufficient so that there is no ambiguity of how the particular results
were obtained.

\subsection{Native Icebergs}

Naturally, a prediction can be done each time the optimization space
gets smaller — after a trade or a new tranche arrival. To match the
case of synthetic icebergs as closely as possible, we decided to evaluate
the prediction results only after each new tranche. The metrics defined
below are calculated across the whole set of icebergs, so we reintroduce
the appropriate indexation.

Let $\hat{v}_{i,r}^{\cdot}$ denote the estimated total size at tranche
$r$ (where ``$\cdot$'' can be any of the ``mean'', ``median''
or ``mode''), $v_{i,r}$ — the actual accumulated volume up to,
but not including, tranche $r$, $p_{i,r}$ — the peak size, and $R_{i}$
— the set of $i$-th iceberg tranches.

\paragraph{Classification}

After a new tranche $r$ arrives, the accumulated volume is $v_{i,r}+p_{i,r}$.
Hence if $v_{i,r}+p_{i,r}<v_{i}$, then the hypothesis is rejected
(the true result is ``negative''); consequently if $v_{i,r}+p_{i,r}<\hat{v}_{i,r}^{\cdot}$,
then the outcome is ``true negative'', otherwise it is ``false
positive''; and vice versa. For $\text{mode}(1),\ldots,\text{mode}(k)$
predictions, consider the prediction true if at least of the them
was true.

\paragraph{Regression}

Compute the residuals $e_{i,r}^{\cdot}=v_{i}-\hat{v}_{i,r}^{\cdot}$,
$r\in R_{i}$. For $\text{mode}(1),\ldots,\text{mode}(k)$ predictions,
select the residual that has a minimum absolute value. Use the residuals
to calculate the standard regression metrics:

\begin{align}
\mathrm{MAE} & =\frac{1}{\left|R\right|}\sum_{i\in C}\sum_{r\in R_{i}}\left|e_{i,r}^{\cdot}\right|,\label{eq:mae}\\
\mathrm{RMSE} & =\sqrt{\frac{1}{\left|R\right|}\sum_{i\in C}\sum_{r\in R_{i}}\left(e_{i,r}^{\cdot}\right)^{2}},\label{eq:rmse}
\end{align}
where $R={\displaystyle \bigcup_{i\in C}}R_{i}$ and $C$ is the set
of complete icebergs.

\subsection{Synthetic Icebergs}

For each complete iceberg with index $i$, the set of test tranches
$R_{i}$ was formed by walking the tranche tree, starting from the
root tranche, visiting each tranche only once and avoiding short chains
(by default, of length less than three). Then the evaluation metrics
were calculated as described below.

Let $\hat{\mathcal{V}}_{i,r}^{\cdot}$ denote the predicted total
size and $\mathcal{V}_{i,r}^{\cdot}$ — the actual accumulated volume
of the $i$-th iceberg up to and including tranche $r$, both quantities
suitably aggregated — e.g. having ``$\text{all}$'', ``$\text{unique}$''
or ``$\text{longest}$'' in place of the ``$\cdot$''.

\paragraph*{Classification}
\begin{itemize}
\item For the last tranche $r_{\max}$, if $\hat{\mathcal{V}}_{i,r_{\max}}^{\cdot}=\mathcal{V}_{i,r_{\max}}^{\cdot}$,
then the case is true positive, otherwise it is a false positive.
\item For all but the last tranche ($r<r_{\max}$), if $\hat{\mathcal{V}}_{i,r}^{\cdot}>\mathcal{V}_{i,r}^{\cdot}$,
then the case is true negative, otherwise it is a false negative.
\end{itemize}

\paragraph{Regression}

Compute the \emph{residuals} $e_{i,r}^{\cdot}=\hat{\mathcal{V}}_{i,r}^{\cdot}-\mathcal{V}_{i,r_{\max}}^{\cdot},\ r\in R_{i}.$
Use equations (\ref{eq:mae}) and (\ref{eq:rmse}) to compute the
regression metrics.

\section{Results}

We estimated $f_{p}(v)$ on one day of ESU19 (E-Mini S\&P 500 futures
contract) FOD LOB log data: from 2019-06-18, roughly 16:45:00 CDT,
to 2019-06-19, 16:00:00 CDT; for synthetic icebergs, $dt$ was set
to 0.3 seconds. The choice of parameters and training intervals is
empirical and may be optimised further, but this falls outside of
the scope of this article. Our evidence suggests that it is reasonable
to include at least one trading session into the learning phase, thus
capturing different order flow regimes throughout the day (see e.g.
\citep[chapter 4]{BouchaudEtAl2018}).

The following figures were produced using the data for the aforementioned
period. For synthetic icebergs, the longest chain volume aggregation
is used.

\subsection{LOB Log Statistics}

Figure~\ref{fig:Action-distribution} summarises the distribution
of actions. Figure~\ref{fig:Trade-volume-distr} shows the distribution
of trade volumes.

\begin{figure}[H]
\centering{}%
\begin{minipage}[t]{0.5\textwidth}%
\begin{center}
\begin{center}
% Created by tikzDevice version 0.12 on 2019-08-28 13:16:45
% !TEX encoding = UTF-8 Unicode
\begin{tikzpicture}[x=1pt,y=1pt]
\definecolor{fillColor}{RGB}{255,255,255}
\path[use as bounding box,fill=fillColor,fill opacity=0.00] (0,0) rectangle (237.16,146.57);
\begin{scope}
\path[clip] (  0.00,  0.00) rectangle (237.16,146.57);
\definecolor{drawColor}{RGB}{255,255,255}
\definecolor{fillColor}{RGB}{255,255,255}

\path[draw=drawColor,line width= 0.5pt,line join=round,line cap=round,fill=fillColor] (  0.00,  0.00) rectangle (237.16,146.57);
\end{scope}
\begin{scope}
\path[clip] ( 38.34, 25.11) rectangle (201.62,142.07);
\definecolor{fillColor}{RGB}{255,255,255}

\path[fill=fillColor] ( 38.34, 25.11) rectangle (201.62,142.07);
\definecolor{drawColor}{gray}{0.92}

\path[draw=drawColor,line width= 0.2pt,line join=round] ( 38.34, 48.05) --
	(201.62, 48.05);

\path[draw=drawColor,line width= 0.2pt,line join=round] ( 38.34, 83.29) --
	(201.62, 83.29);

\path[draw=drawColor,line width= 0.2pt,line join=round] ( 38.34,118.54) --
	(201.62,118.54);

\path[draw=drawColor,line width= 0.5pt,line join=round] ( 38.34, 30.42) --
	(201.62, 30.42);

\path[draw=drawColor,line width= 0.5pt,line join=round] ( 38.34, 65.67) --
	(201.62, 65.67);

\path[draw=drawColor,line width= 0.5pt,line join=round] ( 38.34,100.92) --
	(201.62,100.92);

\path[draw=drawColor,line width= 0.5pt,line join=round] ( 38.34,136.16) --
	(201.62,136.16);

\path[draw=drawColor,line width= 0.5pt,line join=round] ( 57.18, 25.11) --
	( 57.18,142.07);

\path[draw=drawColor,line width= 0.5pt,line join=round] ( 88.58, 25.11) --
	( 88.58,142.07);

\path[draw=drawColor,line width= 0.5pt,line join=round] (119.98, 25.11) --
	(119.98,142.07);

\path[draw=drawColor,line width= 0.5pt,line join=round] (151.38, 25.11) --
	(151.38,142.07);

\path[draw=drawColor,line width= 0.5pt,line join=round] (182.78, 25.11) --
	(182.78,142.07);
\definecolor{fillColor}{gray}{0.35}

\path[fill=fillColor] ( 43.05, 30.42) rectangle ( 71.31, 41.19);

\path[fill=fillColor] ( 74.45, 30.42) rectangle (102.71, 43.46);

\path[fill=fillColor] (105.85, 30.42) rectangle (134.11, 71.62);

\path[fill=fillColor] (137.25, 30.42) rectangle (165.51, 71.76);

\path[fill=fillColor] (168.65, 30.42) rectangle (196.91,136.76);
\definecolor{drawColor}{gray}{0.20}

\path[draw=drawColor,line width= 0.5pt,line join=round,line cap=round] ( 38.34, 25.11) rectangle (201.62,142.07);
\end{scope}
\begin{scope}
\path[clip] (  0.00,  0.00) rectangle (237.16,146.57);
\definecolor{drawColor}{gray}{0.30}

\node[text=drawColor,anchor=base east,inner sep=0pt, outer sep=0pt, scale=  0.72] at ( 34.29, 27.94) {0e+00};

\node[text=drawColor,anchor=base east,inner sep=0pt, outer sep=0pt, scale=  0.72] at ( 34.29, 63.19) {2e+06};

\node[text=drawColor,anchor=base east,inner sep=0pt, outer sep=0pt, scale=  0.72] at ( 34.29, 98.44) {4e+06};

\node[text=drawColor,anchor=base east,inner sep=0pt, outer sep=0pt, scale=  0.72] at ( 34.29,133.68) {6e+06};
\end{scope}
\begin{scope}
\path[clip] (  0.00,  0.00) rectangle (237.16,146.57);
\definecolor{drawColor}{gray}{0.20}

\path[draw=drawColor,line width= 0.5pt,line join=round] ( 36.09, 30.42) --
	( 38.34, 30.42);

\path[draw=drawColor,line width= 0.5pt,line join=round] ( 36.09, 65.67) --
	( 38.34, 65.67);

\path[draw=drawColor,line width= 0.5pt,line join=round] ( 36.09,100.92) --
	( 38.34,100.92);

\path[draw=drawColor,line width= 0.5pt,line join=round] ( 36.09,136.16) --
	( 38.34,136.16);
\end{scope}
\begin{scope}
\path[clip] (  0.00,  0.00) rectangle (237.16,146.57);
\definecolor{drawColor}{gray}{0.20}

\path[draw=drawColor,line width= 0.5pt,line join=round] (201.62, 30.37) --
	(203.87, 30.37);

\path[draw=drawColor,line width= 0.5pt,line join=round] (201.62, 56.92) --
	(203.87, 56.92);

\path[draw=drawColor,line width= 0.5pt,line join=round] (201.62, 83.47) --
	(203.87, 83.47);

\path[draw=drawColor,line width= 0.5pt,line join=round] (201.62,110.26) --
	(203.87,110.26);

\path[draw=drawColor,line width= 0.5pt,line join=round] (201.62,136.81) --
	(203.87,136.81);
\end{scope}
\begin{scope}
\path[clip] (  0.00,  0.00) rectangle (237.16,146.57);
\definecolor{drawColor}{gray}{0.30}

\node[text=drawColor,anchor=base west,inner sep=0pt, outer sep=0pt, scale=  0.72] at (205.67, 27.89) {0\%};

\node[text=drawColor,anchor=base west,inner sep=0pt, outer sep=0pt, scale=  0.72] at (205.67, 54.44) {25\%};

\node[text=drawColor,anchor=base west,inner sep=0pt, outer sep=0pt, scale=  0.72] at (205.67, 80.99) {50\%};

\node[text=drawColor,anchor=base west,inner sep=0pt, outer sep=0pt, scale=  0.72] at (205.67,107.78) {75\%};

\node[text=drawColor,anchor=base west,inner sep=0pt, outer sep=0pt, scale=  0.72] at (205.67,134.33) {100\%};
\end{scope}
\begin{scope}
\path[clip] (  0.00,  0.00) rectangle (237.16,146.57);
\definecolor{drawColor}{gray}{0.20}

\path[draw=drawColor,line width= 0.5pt,line join=round] ( 57.18, 22.86) --
	( 57.18, 25.11);

\path[draw=drawColor,line width= 0.5pt,line join=round] ( 88.58, 22.86) --
	( 88.58, 25.11);

\path[draw=drawColor,line width= 0.5pt,line join=round] (119.98, 22.86) --
	(119.98, 25.11);

\path[draw=drawColor,line width= 0.5pt,line join=round] (151.38, 22.86) --
	(151.38, 25.11);

\path[draw=drawColor,line width= 0.5pt,line join=round] (182.78, 22.86) --
	(182.78, 25.11);
\end{scope}
\begin{scope}
\path[clip] (  0.00,  0.00) rectangle (237.16,146.57);
\definecolor{drawColor}{gray}{0.30}

\node[text=drawColor,anchor=base,inner sep=0pt, outer sep=0pt, scale=  0.72] at ( 57.18, 16.10) {TRADE};

\node[text=drawColor,anchor=base,inner sep=0pt, outer sep=0pt, scale=  0.72] at ( 88.58, 16.10) {MODIFY};

\node[text=drawColor,anchor=base,inner sep=0pt, outer sep=0pt, scale=  0.72] at (119.98, 16.10) {DELETE};

\node[text=drawColor,anchor=base,inner sep=0pt, outer sep=0pt, scale=  0.72] at (151.38, 16.10) {LIMIT};

\node[text=drawColor,anchor=base,inner sep=0pt, outer sep=0pt, scale=  0.72] at (182.78, 16.10) {(ALL)};
\end{scope}
\begin{scope}
\path[clip] (  0.00,  0.00) rectangle (237.16,146.57);
\definecolor{drawColor}{RGB}{0,0,0}

\node[text=drawColor,anchor=base,inner sep=0pt, outer sep=0pt, scale=  0.90] at (119.98,  6.25) {Action};
\end{scope}
\begin{scope}
\path[clip] (  0.00,  0.00) rectangle (237.16,146.57);
\definecolor{drawColor}{RGB}{0,0,0}

\node[text=drawColor,rotate= 90.00,anchor=base,inner sep=0pt, outer sep=0pt, scale=  0.90] at ( 10.70, 83.59) {Action count};
\end{scope}
\begin{scope}
\path[clip] (  0.00,  0.00) rectangle (237.16,146.57);
\definecolor{drawColor}{RGB}{0,0,0}

\node[text=drawColor,rotate=-90.00,anchor=base,inner sep=0pt, outer sep=0pt, scale=  0.90] at (224.71, 83.59) {\% of all log records};
\end{scope}
\end{tikzpicture}
\par\end{center}\caption{\label{fig:Action-distribution}Action distribution.}
\par\end{center}%
\end{minipage}\hfill{}%
\begin{minipage}[t]{0.5\textwidth}%
\begin{center}
\begin{center}
% Created by tikzDevice version 0.12 on 2019-08-28 13:21:27
% !TEX encoding = UTF-8 Unicode
\begin{tikzpicture}[x=1pt,y=1pt]
\definecolor{fillColor}{RGB}{255,255,255}
\path[use as bounding box,fill=fillColor,fill opacity=0.00] (0,0) rectangle (237.16,146.57);
\begin{scope}
\path[clip] (  0.00,  0.00) rectangle (237.16,146.57);
\definecolor{drawColor}{RGB}{255,255,255}
\definecolor{fillColor}{RGB}{255,255,255}

\path[draw=drawColor,line width= 0.5pt,line join=round,line cap=round,fill=fillColor] (  0.00,  0.00) rectangle (237.16,146.57);
\end{scope}
\begin{scope}
\path[clip] ( 38.34, 25.11) rectangle (232.66,142.07);
\definecolor{fillColor}{RGB}{255,255,255}

\path[fill=fillColor] ( 38.34, 25.11) rectangle (232.66,142.07);
\definecolor{drawColor}{gray}{0.92}

\path[draw=drawColor,line width= 0.2pt,line join=round] ( 38.34, 30.55) --
	(232.66, 30.55);

\path[draw=drawColor,line width= 0.2pt,line join=round] ( 38.34, 71.07) --
	(232.66, 71.07);

\path[draw=drawColor,line width= 0.2pt,line join=round] ( 38.34,111.63) --
	(232.66,111.63);

\path[draw=drawColor,line width= 0.2pt,line join=round] ( 74.29, 25.11) --
	( 74.29,142.07);

\path[draw=drawColor,line width= 0.2pt,line join=round] (124.98, 25.11) --
	(124.98,142.07);

\path[draw=drawColor,line width= 0.2pt,line join=round] (175.68, 25.11) --
	(175.68,142.07);

\path[draw=drawColor,line width= 0.2pt,line join=round] (226.37, 25.11) --
	(226.37,142.07);

\path[draw=drawColor,line width= 0.5pt,line join=round] ( 38.34, 50.81) --
	(232.66, 50.81);

\path[draw=drawColor,line width= 0.5pt,line join=round] ( 38.34, 91.33) --
	(232.66, 91.33);

\path[draw=drawColor,line width= 0.5pt,line join=round] ( 38.34,131.93) --
	(232.66,131.93);

\path[draw=drawColor,line width= 0.5pt,line join=round] ( 48.94, 25.11) --
	( 48.94,142.07);

\path[draw=drawColor,line width= 0.5pt,line join=round] ( 99.64, 25.11) --
	( 99.64,142.07);

\path[draw=drawColor,line width= 0.5pt,line join=round] (150.33, 25.11) --
	(150.33,142.07);

\path[draw=drawColor,line width= 0.5pt,line join=round] (201.02, 25.11) --
	(201.02,142.07);
\definecolor{fillColor}{gray}{0.35}

\path[fill=fillColor] ( 47.18, 30.42) rectangle ( 50.71,136.76);

\path[fill=fillColor] ( 50.71, 30.42) rectangle ( 54.24,121.45);

\path[fill=fillColor] ( 54.24, 30.42) rectangle ( 57.77,106.69);

\path[fill=fillColor] ( 57.77, 30.42) rectangle ( 61.31, 98.93);

\path[fill=fillColor] ( 61.31, 30.42) rectangle ( 64.84, 95.07);

\path[fill=fillColor] ( 64.84, 30.42) rectangle ( 68.37, 88.47);

\path[fill=fillColor] ( 68.37, 30.42) rectangle ( 71.91, 84.19);

\path[fill=fillColor] ( 71.91, 30.42) rectangle ( 75.44, 85.34);

\path[fill=fillColor] ( 75.44, 30.42) rectangle ( 78.97, 76.64);

\path[fill=fillColor] ( 78.97, 30.42) rectangle ( 82.51, 73.48);

\path[fill=fillColor] ( 82.51, 30.42) rectangle ( 86.04, 69.80);

\path[fill=fillColor] ( 86.04, 30.42) rectangle ( 89.57, 70.10);

\path[fill=fillColor] ( 89.57, 30.42) rectangle ( 93.11, 63.38);

\path[fill=fillColor] ( 93.11, 30.42) rectangle ( 96.64, 63.99);

\path[fill=fillColor] ( 96.64, 30.42) rectangle (100.17, 65.76);

\path[fill=fillColor] (100.17, 30.42) rectangle (103.70, 55.94);

\path[fill=fillColor] (103.70, 30.42) rectangle (107.24, 56.41);

\path[fill=fillColor] (107.24, 30.42) rectangle (110.77, 51.64);

\path[fill=fillColor] (110.77, 30.42) rectangle (114.30, 49.90);

\path[fill=fillColor] (114.30, 30.42) rectangle (117.84, 48.89);

\path[fill=fillColor] (117.84, 30.42) rectangle (121.37, 43.15);

\path[fill=fillColor] (121.37, 30.42) rectangle (124.90, 52.39);

\path[fill=fillColor] (124.90, 30.42) rectangle (128.44, 49.90);

\path[fill=fillColor] (128.44, 30.42) rectangle (131.97, 40.96);

\path[fill=fillColor] (131.97, 30.42) rectangle (135.50, 49.90);

\path[fill=fillColor] (135.50, 30.42) rectangle (139.03, 40.96);

\path[fill=fillColor] (139.03, 30.42) rectangle (142.57, 38.19);

\path[fill=fillColor] (142.57, 30.42) rectangle (146.10, 43.15);

\path[fill=fillColor] (146.10, 30.42) rectangle (149.63, 38.19);

\path[fill=fillColor] (149.63, 30.42) rectangle (153.17, 34.67);

\path[fill=fillColor] (153.17, 30.42) rectangle (156.70, 34.67);

\path[fill=fillColor] (156.70, 30.42) rectangle (160.23, 30.42);

\path[fill=fillColor] (160.23, 30.42) rectangle (163.77, 30.42);

\path[fill=fillColor] (163.77, 30.42) rectangle (167.30, 30.42);

\path[fill=fillColor] (167.30, 30.42) rectangle (170.83, 30.42);

\path[fill=fillColor] (170.83, 30.42) rectangle (174.36, 30.42);

\path[fill=fillColor] (174.36, 30.42) rectangle (177.90, 34.67);

\path[fill=fillColor] (177.90, 30.42) rectangle (181.43, 30.42);

\path[fill=fillColor] (181.43, 30.42) rectangle (184.96, 30.42);

\path[fill=fillColor] (184.96, 30.42) rectangle (188.50, 30.42);

\path[fill=fillColor] (188.50, 30.42) rectangle (192.03, 30.42);

\path[fill=fillColor] (192.03, 30.42) rectangle (195.56, 30.42);

\path[fill=fillColor] (195.56, 30.42) rectangle (199.10, 30.42);

\path[fill=fillColor] (199.10, 30.42) rectangle (202.63, 30.42);

\path[fill=fillColor] (202.63, 30.42) rectangle (206.16, 30.42);

\path[fill=fillColor] (206.16, 30.42) rectangle (209.70, 30.42);

\path[fill=fillColor] (209.70, 30.42) rectangle (213.23, 30.42);

\path[fill=fillColor] (213.23, 30.42) rectangle (216.76, 30.42);

\path[fill=fillColor] (216.76, 30.42) rectangle (220.29, 34.67);

\path[fill=fillColor] (220.29, 30.42) rectangle (223.83, 34.67);
\definecolor{drawColor}{gray}{0.20}

\path[draw=drawColor,line width= 0.5pt,line join=round,line cap=round] ( 38.34, 25.11) rectangle (232.66,142.07);
\end{scope}
\begin{scope}
\path[clip] (  0.00,  0.00) rectangle (237.16,146.57);
\definecolor{drawColor}{gray}{0.30}

\node[text=drawColor,anchor=base east,inner sep=0pt, outer sep=0pt, scale=  0.72] at ( 34.29, 48.33) {1e+01};

\node[text=drawColor,anchor=base east,inner sep=0pt, outer sep=0pt, scale=  0.72] at ( 34.29, 88.85) {1e+03};

\node[text=drawColor,anchor=base east,inner sep=0pt, outer sep=0pt, scale=  0.72] at ( 34.29,129.46) {1e+05};
\end{scope}
\begin{scope}
\path[clip] (  0.00,  0.00) rectangle (237.16,146.57);
\definecolor{drawColor}{gray}{0.20}

\path[draw=drawColor,line width= 0.5pt,line join=round] ( 36.09, 50.81) --
	( 38.34, 50.81);

\path[draw=drawColor,line width= 0.5pt,line join=round] ( 36.09, 91.33) --
	( 38.34, 91.33);

\path[draw=drawColor,line width= 0.5pt,line join=round] ( 36.09,131.93) --
	( 38.34,131.93);
\end{scope}
\begin{scope}
\path[clip] (  0.00,  0.00) rectangle (237.16,146.57);
\definecolor{drawColor}{gray}{0.20}

\path[draw=drawColor,line width= 0.5pt,line join=round] ( 48.94, 22.86) --
	( 48.94, 25.11);

\path[draw=drawColor,line width= 0.5pt,line join=round] ( 99.64, 22.86) --
	( 99.64, 25.11);

\path[draw=drawColor,line width= 0.5pt,line join=round] (150.33, 22.86) --
	(150.33, 25.11);

\path[draw=drawColor,line width= 0.5pt,line join=round] (201.02, 22.86) --
	(201.02, 25.11);
\end{scope}
\begin{scope}
\path[clip] (  0.00,  0.00) rectangle (237.16,146.57);
\definecolor{drawColor}{gray}{0.30}

\node[text=drawColor,anchor=base,inner sep=0pt, outer sep=0pt, scale=  0.72] at ( 48.94, 16.10) {0};

\node[text=drawColor,anchor=base,inner sep=0pt, outer sep=0pt, scale=  0.72] at ( 99.64, 16.10) {200};

\node[text=drawColor,anchor=base,inner sep=0pt, outer sep=0pt, scale=  0.72] at (150.33, 16.10) {400};

\node[text=drawColor,anchor=base,inner sep=0pt, outer sep=0pt, scale=  0.72] at (201.02, 16.10) {600};
\end{scope}
\begin{scope}
\path[clip] (  0.00,  0.00) rectangle (237.16,146.57);
\definecolor{drawColor}{RGB}{0,0,0}

\node[text=drawColor,anchor=base,inner sep=0pt, outer sep=0pt, scale=  0.90] at (135.50,  6.25) {Volume};
\end{scope}
\begin{scope}
\path[clip] (  0.00,  0.00) rectangle (237.16,146.57);
\definecolor{drawColor}{RGB}{0,0,0}

\node[text=drawColor,rotate= 90.00,anchor=base,inner sep=0pt, outer sep=0pt, scale=  0.90] at ( 10.70, 83.59) {Trade count};
\end{scope}
\end{tikzpicture}
\par\end{center}\caption{\label{fig:Trade-volume-distr}Trade volume distribution.}
\par\end{center}%
\end{minipage}
\end{figure}

\subsection{Detection Results}

\begin{figure}[h]
  \hfill{}
  \subfigure[Synthetic icebergs]{% Created by tikzDevice version 0.12 on 2019-08-28 13:16:05
% !TEX encoding = UTF-8 Unicode
\begin{tikzpicture}[x=1pt,y=1pt]
\definecolor{fillColor}{RGB}{255,255,255}
\path[use as bounding box,fill=fillColor,fill opacity=0.00] (0,0) rectangle (217.80,134.61);
\begin{scope}
\path[clip] (  0.00,  0.00) rectangle (217.80,134.61);
\definecolor{drawColor}{RGB}{255,255,255}
\definecolor{fillColor}{RGB}{255,255,255}

\path[draw=drawColor,line width= 0.5pt,line join=round,line cap=round,fill=fillColor] (  0.00,  0.00) rectangle (217.80,134.61);
\end{scope}
\begin{scope}
\path[clip] ( 36.74, 25.11) rectangle (182.26,130.11);
\definecolor{fillColor}{RGB}{255,255,255}

\path[fill=fillColor] ( 36.74, 25.11) rectangle (182.26,130.11);
\definecolor{drawColor}{gray}{0.92}

\path[draw=drawColor,line width= 0.2pt,line join=round] ( 36.74, 43.44) --
	(182.26, 43.44);

\path[draw=drawColor,line width= 0.2pt,line join=round] ( 36.74, 70.56) --
	(182.26, 70.56);

\path[draw=drawColor,line width= 0.2pt,line join=round] ( 36.74, 97.69) --
	(182.26, 97.69);

\path[draw=drawColor,line width= 0.2pt,line join=round] ( 36.74,124.81) --
	(182.26,124.81);

\path[draw=drawColor,line width= 0.5pt,line join=round] ( 36.74, 29.88) --
	(182.26, 29.88);

\path[draw=drawColor,line width= 0.5pt,line join=round] ( 36.74, 57.00) --
	(182.26, 57.00);

\path[draw=drawColor,line width= 0.5pt,line join=round] ( 36.74, 84.13) --
	(182.26, 84.13);

\path[draw=drawColor,line width= 0.5pt,line join=round] ( 36.74,111.25) --
	(182.26,111.25);

\path[draw=drawColor,line width= 0.5pt,line join=round] ( 64.03, 25.11) --
	( 64.03,130.11);

\path[draw=drawColor,line width= 0.5pt,line join=round] (109.50, 25.11) --
	(109.50,130.11);

\path[draw=drawColor,line width= 0.5pt,line join=round] (154.97, 25.11) --
	(154.97,130.11);
\definecolor{fillColor}{gray}{0.35}

\path[fill=fillColor] ( 43.56, 29.88) rectangle ( 84.49,125.34);

\path[fill=fillColor] ( 89.04, 29.88) rectangle (129.96, 97.25);

\path[fill=fillColor] (134.51, 29.88) rectangle (175.44, 57.96);
\definecolor{drawColor}{gray}{0.20}

\path[draw=drawColor,line width= 0.5pt,line join=round,line cap=round] ( 36.74, 25.11) rectangle (182.26,130.11);
\end{scope}
\begin{scope}
\path[clip] (  0.00,  0.00) rectangle (217.80,134.61);
\definecolor{drawColor}{gray}{0.30}

\node[text=drawColor,anchor=base east,inner sep=0pt, outer sep=0pt, scale=  0.72] at ( 32.69, 27.40) {0};

\node[text=drawColor,anchor=base east,inner sep=0pt, outer sep=0pt, scale=  0.72] at ( 32.69, 54.52) {20000};

\node[text=drawColor,anchor=base east,inner sep=0pt, outer sep=0pt, scale=  0.72] at ( 32.69, 81.65) {40000};

\node[text=drawColor,anchor=base east,inner sep=0pt, outer sep=0pt, scale=  0.72] at ( 32.69,108.77) {60000};
\end{scope}
\begin{scope}
\path[clip] (  0.00,  0.00) rectangle (217.80,134.61);
\definecolor{drawColor}{gray}{0.20}

\path[draw=drawColor,line width= 0.5pt,line join=round] ( 34.49, 29.88) --
	( 36.74, 29.88);

\path[draw=drawColor,line width= 0.5pt,line join=round] ( 34.49, 57.00) --
	( 36.74, 57.00);

\path[draw=drawColor,line width= 0.5pt,line join=round] ( 34.49, 84.13) --
	( 36.74, 84.13);

\path[draw=drawColor,line width= 0.5pt,line join=round] ( 34.49,111.25) --
	( 36.74,111.25);
\end{scope}
\begin{scope}
\path[clip] (  0.00,  0.00) rectangle (217.80,134.61);
\definecolor{drawColor}{gray}{0.20}

\path[draw=drawColor,line width= 0.5pt,line join=round] (182.26, 29.83) --
	(184.51, 29.83);

\path[draw=drawColor,line width= 0.5pt,line join=round] (182.26, 53.67) --
	(184.51, 53.67);

\path[draw=drawColor,line width= 0.5pt,line join=round] (182.26, 77.50) --
	(184.51, 77.50);

\path[draw=drawColor,line width= 0.5pt,line join=round] (182.26,101.55) --
	(184.51,101.55);

\path[draw=drawColor,line width= 0.5pt,line join=round] (182.26,125.38) --
	(184.51,125.38);
\end{scope}
\begin{scope}
\path[clip] (  0.00,  0.00) rectangle (217.80,134.61);
\definecolor{drawColor}{gray}{0.30}

\node[text=drawColor,anchor=base west,inner sep=0pt, outer sep=0pt, scale=  0.72] at (186.31, 27.35) {0\%};

\node[text=drawColor,anchor=base west,inner sep=0pt, outer sep=0pt, scale=  0.72] at (186.31, 51.19) {25\%};

\node[text=drawColor,anchor=base west,inner sep=0pt, outer sep=0pt, scale=  0.72] at (186.31, 75.02) {50\%};

\node[text=drawColor,anchor=base west,inner sep=0pt, outer sep=0pt, scale=  0.72] at (186.31, 99.07) {75\%};

\node[text=drawColor,anchor=base west,inner sep=0pt, outer sep=0pt, scale=  0.72] at (186.31,122.90) {100\%};
\end{scope}
\begin{scope}
\path[clip] (  0.00,  0.00) rectangle (217.80,134.61);
\definecolor{drawColor}{gray}{0.20}

\path[draw=drawColor,line width= 0.5pt,line join=round] ( 64.03, 22.86) --
	( 64.03, 25.11);

\path[draw=drawColor,line width= 0.5pt,line join=round] (109.50, 22.86) --
	(109.50, 25.11);

\path[draw=drawColor,line width= 0.5pt,line join=round] (154.97, 22.86) --
	(154.97, 25.11);
\end{scope}
\begin{scope}
\path[clip] (  0.00,  0.00) rectangle (217.80,134.61);
\definecolor{drawColor}{gray}{0.30}

\node[text=drawColor,anchor=base,inner sep=0pt, outer sep=0pt, scale=  0.72] at ( 64.03, 16.10) {all};

\node[text=drawColor,anchor=base,inner sep=0pt, outer sep=0pt, scale=  0.72] at (109.50, 16.10) {cancelled};

\node[text=drawColor,anchor=base,inner sep=0pt, outer sep=0pt, scale=  0.72] at (154.97, 16.10) {complete};
\end{scope}
\begin{scope}
\path[clip] (  0.00,  0.00) rectangle (217.80,134.61);
\definecolor{drawColor}{RGB}{0,0,0}

\node[text=drawColor,anchor=base,inner sep=0pt, outer sep=0pt, scale=  0.90] at (109.50,  6.25) {Completion};
\end{scope}
\begin{scope}
\path[clip] (  0.00,  0.00) rectangle (217.80,134.61);
\definecolor{drawColor}{RGB}{0,0,0}

\node[text=drawColor,rotate= 90.00,anchor=base,inner sep=0pt, outer sep=0pt, scale=  0.90] at ( 10.70, 77.61) {Iceberg count};
\end{scope}
\begin{scope}
\path[clip] (  0.00,  0.00) rectangle (217.80,134.61);
\definecolor{drawColor}{RGB}{0,0,0}

\node[text=drawColor,rotate=-90.00,anchor=base,inner sep=0pt, outer sep=0pt, scale=  0.90] at (205.35, 77.61) {\% of all icebergs};
\end{scope}
\end{tikzpicture}}
  \hfill{}
  \subfigure[Native icebergs]{% Created by tikzDevice version 0.12 on 2019-08-28 13:16:22
% !TEX encoding = UTF-8 Unicode
\begin{tikzpicture}[x=1pt,y=1pt]
\definecolor{fillColor}{RGB}{255,255,255}
\path[use as bounding box,fill=fillColor,fill opacity=0.00] (0,0) rectangle (217.80,134.61);
\begin{scope}
\path[clip] (  0.00,  0.00) rectangle (217.80,134.61);
\definecolor{drawColor}{RGB}{255,255,255}
\definecolor{fillColor}{RGB}{255,255,255}

\path[draw=drawColor,line width= 0.5pt,line join=round,line cap=round,fill=fillColor] (  0.00,  0.00) rectangle (217.80,134.61);
\end{scope}
\begin{scope}
\path[clip] ( 33.14, 25.11) rectangle (182.26,130.11);
\definecolor{fillColor}{RGB}{255,255,255}

\path[fill=fillColor] ( 33.14, 25.11) rectangle (182.26,130.11);
\definecolor{drawColor}{gray}{0.92}

\path[draw=drawColor,line width= 0.2pt,line join=round] ( 33.14, 43.87) --
	(182.26, 43.87);

\path[draw=drawColor,line width= 0.2pt,line join=round] ( 33.14, 71.84) --
	(182.26, 71.84);

\path[draw=drawColor,line width= 0.2pt,line join=round] ( 33.14, 99.82) --
	(182.26, 99.82);

\path[draw=drawColor,line width= 0.2pt,line join=round] ( 33.14,127.80) --
	(182.26,127.80);

\path[draw=drawColor,line width= 0.5pt,line join=round] ( 33.14, 29.88) --
	(182.26, 29.88);

\path[draw=drawColor,line width= 0.5pt,line join=round] ( 33.14, 57.86) --
	(182.26, 57.86);

\path[draw=drawColor,line width= 0.5pt,line join=round] ( 33.14, 85.83) --
	(182.26, 85.83);

\path[draw=drawColor,line width= 0.5pt,line join=round] ( 33.14,113.81) --
	(182.26,113.81);

\path[draw=drawColor,line width= 0.5pt,line join=round] ( 61.10, 25.11) --
	( 61.10,130.11);

\path[draw=drawColor,line width= 0.5pt,line join=round] (107.70, 25.11) --
	(107.70,130.11);

\path[draw=drawColor,line width= 0.5pt,line join=round] (154.30, 25.11) --
	(154.30,130.11);
\definecolor{fillColor}{gray}{0.35}

\path[fill=fillColor] ( 40.13, 29.88) rectangle ( 82.07,125.34);

\path[fill=fillColor] ( 86.73, 29.88) rectangle (128.67, 36.31);

\path[fill=fillColor] (133.33, 29.88) rectangle (175.27,118.90);
\definecolor{drawColor}{gray}{0.20}

\path[draw=drawColor,line width= 0.5pt,line join=round,line cap=round] ( 33.14, 25.11) rectangle (182.26,130.11);
\end{scope}
\begin{scope}
\path[clip] (  0.00,  0.00) rectangle (217.80,134.61);
\definecolor{drawColor}{gray}{0.30}

\node[text=drawColor,anchor=base east,inner sep=0pt, outer sep=0pt, scale=  0.72] at ( 29.09, 27.40) {0};

\node[text=drawColor,anchor=base east,inner sep=0pt, outer sep=0pt, scale=  0.72] at ( 29.09, 55.38) {500};

\node[text=drawColor,anchor=base east,inner sep=0pt, outer sep=0pt, scale=  0.72] at ( 29.09, 83.35) {1000};

\node[text=drawColor,anchor=base east,inner sep=0pt, outer sep=0pt, scale=  0.72] at ( 29.09,111.33) {1500};
\end{scope}
\begin{scope}
\path[clip] (  0.00,  0.00) rectangle (217.80,134.61);
\definecolor{drawColor}{gray}{0.20}

\path[draw=drawColor,line width= 0.5pt,line join=round] ( 30.89, 29.88) --
	( 33.14, 29.88);

\path[draw=drawColor,line width= 0.5pt,line join=round] ( 30.89, 57.86) --
	( 33.14, 57.86);

\path[draw=drawColor,line width= 0.5pt,line join=round] ( 30.89, 85.83) --
	( 33.14, 85.83);

\path[draw=drawColor,line width= 0.5pt,line join=round] ( 30.89,113.81) --
	( 33.14,113.81);
\end{scope}
\begin{scope}
\path[clip] (  0.00,  0.00) rectangle (217.80,134.61);
\definecolor{drawColor}{gray}{0.20}

\path[draw=drawColor,line width= 0.5pt,line join=round] (182.26, 29.83) --
	(184.51, 29.83);

\path[draw=drawColor,line width= 0.5pt,line join=round] (182.26, 53.67) --
	(184.51, 53.67);

\path[draw=drawColor,line width= 0.5pt,line join=round] (182.26, 77.50) --
	(184.51, 77.50);

\path[draw=drawColor,line width= 0.5pt,line join=round] (182.26,101.55) --
	(184.51,101.55);

\path[draw=drawColor,line width= 0.5pt,line join=round] (182.26,125.38) --
	(184.51,125.38);
\end{scope}
\begin{scope}
\path[clip] (  0.00,  0.00) rectangle (217.80,134.61);
\definecolor{drawColor}{gray}{0.30}

\node[text=drawColor,anchor=base west,inner sep=0pt, outer sep=0pt, scale=  0.72] at (186.31, 27.35) {0\%};

\node[text=drawColor,anchor=base west,inner sep=0pt, outer sep=0pt, scale=  0.72] at (186.31, 51.19) {25\%};

\node[text=drawColor,anchor=base west,inner sep=0pt, outer sep=0pt, scale=  0.72] at (186.31, 75.02) {50\%};

\node[text=drawColor,anchor=base west,inner sep=0pt, outer sep=0pt, scale=  0.72] at (186.31, 99.07) {75\%};

\node[text=drawColor,anchor=base west,inner sep=0pt, outer sep=0pt, scale=  0.72] at (186.31,122.90) {100\%};
\end{scope}
\begin{scope}
\path[clip] (  0.00,  0.00) rectangle (217.80,134.61);
\definecolor{drawColor}{gray}{0.20}

\path[draw=drawColor,line width= 0.5pt,line join=round] ( 61.10, 22.86) --
	( 61.10, 25.11);

\path[draw=drawColor,line width= 0.5pt,line join=round] (107.70, 22.86) --
	(107.70, 25.11);

\path[draw=drawColor,line width= 0.5pt,line join=round] (154.30, 22.86) --
	(154.30, 25.11);
\end{scope}
\begin{scope}
\path[clip] (  0.00,  0.00) rectangle (217.80,134.61);
\definecolor{drawColor}{gray}{0.30}

\node[text=drawColor,anchor=base,inner sep=0pt, outer sep=0pt, scale=  0.72] at ( 61.10, 16.10) {all};

\node[text=drawColor,anchor=base,inner sep=0pt, outer sep=0pt, scale=  0.72] at (107.70, 16.10) {cancelled};

\node[text=drawColor,anchor=base,inner sep=0pt, outer sep=0pt, scale=  0.72] at (154.30, 16.10) {complete};
\end{scope}
\begin{scope}
\path[clip] (  0.00,  0.00) rectangle (217.80,134.61);
\definecolor{drawColor}{RGB}{0,0,0}

\node[text=drawColor,anchor=base,inner sep=0pt, outer sep=0pt, scale=  0.90] at (107.70,  6.25) {Completion};
\end{scope}
\begin{scope}
\path[clip] (  0.00,  0.00) rectangle (217.80,134.61);
\definecolor{drawColor}{RGB}{0,0,0}

\node[text=drawColor,rotate= 90.00,anchor=base,inner sep=0pt, outer sep=0pt, scale=  0.90] at ( 10.70, 77.61) {Iceberg count};
\end{scope}
\begin{scope}
\path[clip] (  0.00,  0.00) rectangle (217.80,134.61);
\definecolor{drawColor}{RGB}{0,0,0}

\node[text=drawColor,rotate=-90.00,anchor=base,inner sep=0pt, outer sep=0pt, scale=  0.90] at (205.35, 77.61) {\% of all icebergs};
\end{scope}
\end{tikzpicture}}
  \hfill{}
  \caption{Iceberg completion state distribution on ESM19 (E-Mini S\&P 500) futures FOD LOB log data (2019\=/06\=/18~16:45:00~CDT – 2019\=/06\=/19~16:00:00~CDT).}
  \label{fig:Iceberg-completion-dist}
\end{figure}
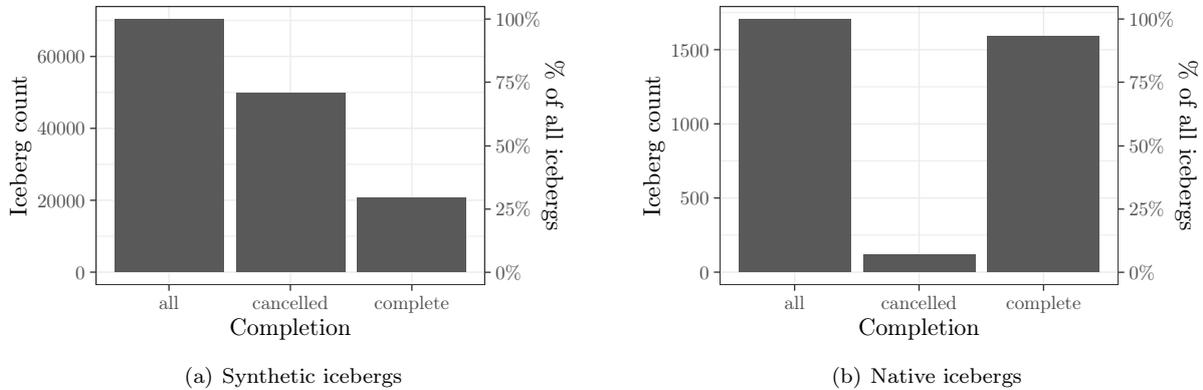

Fig. \ref{fig:Iceberg-completion-dist} shows the proportion of completed
and cancelled icebergs of both types.

The proportion of iceberg orders to all orders on one trading day
is shown in fig.~\ref{fig:Iceberg-proportions} in terms of both
volume and number of orders. In case of synthetic icebergs, the results
depend on the minimum number of tranches per iceberg — that is, the
number of tranches after which their sequence is considered an iceberg.
By increasing this parameter, we decrease the false positive rate
at the cost of disregarding all icebergs of shorter lengths.

We divide the total volume of all iceberg orders by the total traded
volume of all orders (like e.g. \citep{Frey2009} do), and not the
total daily limit order volume. This ratio makes more sense because
only executed icebergs can be detected, which surely constitute only
a fraction of all resting hidden volume. We estimate that 4\% of all
traded volume is contributed by native icebergs, while the volume
contributed by synthetic icebergs ranges from 3.3 to 14.3\%, depending
on the minimum number of tranches. This is in agreement with some
of the results reported in the literature as alluded to earlier in
section \ref{sec:Existing-Literature-Overview}.

Moreover, as \citep{FlemingMizrach2008} note, usually there is no
hidden depth, but when it is present, it is substantial. This is especially
true for native icebergs, that constitute 0.06\% of all orders by
number, but 4\% by volume; see fig.~\ref{fig:Iceberg-proportions}.

\begin{figure}[H]
  \hfill{}
  \subfigure[Density]{\input{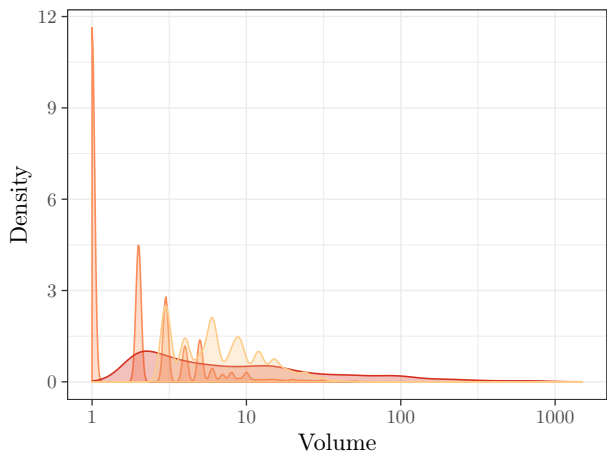}}
  \hfill{}
  \subfigure[Individual probabilites (cutoff at 100 units)]{\input{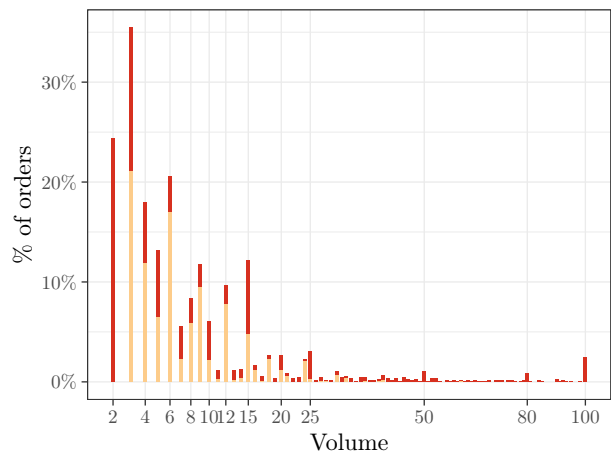}}
  \hfill{}
  \caption{Order size distribution.}
  \label{fig:Order-volume-distribution}
\end{figure}

In addition, the following size-related distributions are estimated:
\begin{itemize}
\item Trade volume (fig.~\ref{fig:Order-volume-distribution}). At least
with native icebergs, we confirm the finding of \citep{Christensen2013}
that order sizes to be multiples of 5, like 15, 25, 50 or 100 as can
be seen in the right panel — this might be inidicative of a human
bias.
\item Peak volume (fig.~\ref{fig:Peak-volume-dist}).
\item Number of tranches per iceberg (fig.~\ref{fig:Tranches-count}).
\end{itemize}
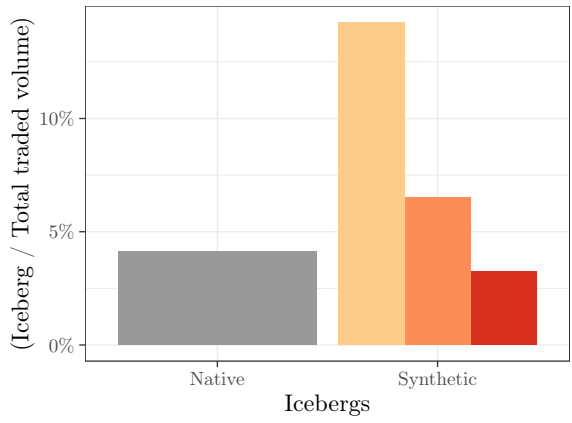
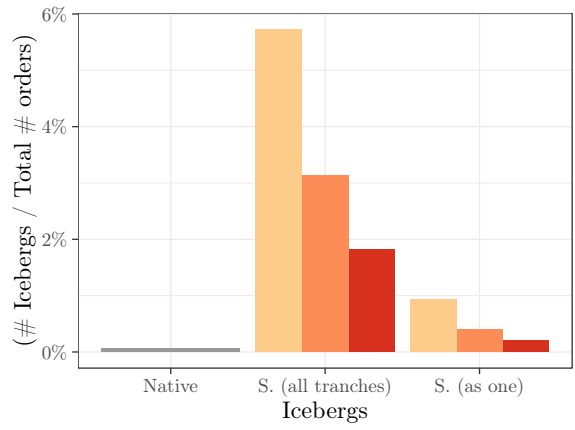
\begin{figure}[H]
  \hfill{}
  \subfigure[By volume]{% Created by tikzDevice version 0.12 on 2019-08-28 16:05:36
% !TEX encoding = UTF-8 Unicode
\begin{tikzpicture}[x=1pt,y=1pt]
\definecolor{fillColor}{RGB}{255,255,255}
\path[use as bounding box,fill=fillColor,fill opacity=0.00] (0,0) rectangle (217.80,196.02);
\begin{scope}
\path[clip] (  0.00,  0.00) rectangle (217.80,196.02);
\definecolor{drawColor}{RGB}{255,255,255}
\definecolor{fillColor}{RGB}{255,255,255}

\path[draw=drawColor,line width= 0.5pt,line join=round,line cap=round,fill=fillColor] (  0.00,  0.00) rectangle (217.80,196.02);
\end{scope}
\begin{scope}
\path[clip] ( 31.94, 57.56) rectangle (213.30,191.52);
\definecolor{fillColor}{RGB}{255,255,255}

\path[fill=fillColor] ( 31.94, 57.56) rectangle (213.30,191.52);
\definecolor{drawColor}{gray}{0.92}

\path[draw=drawColor,line width= 0.2pt,line join=round] ( 31.94, 85.01) --
	(213.30, 85.01);

\path[draw=drawColor,line width= 0.2pt,line join=round] ( 31.94,127.72) --
	(213.30,127.72);

\path[draw=drawColor,line width= 0.2pt,line join=round] ( 31.94,170.44) --
	(213.30,170.44);

\path[draw=drawColor,line width= 0.5pt,line join=round] ( 31.94, 63.65) --
	(213.30, 63.65);

\path[draw=drawColor,line width= 0.5pt,line join=round] ( 31.94,106.37) --
	(213.30,106.37);

\path[draw=drawColor,line width= 0.5pt,line join=round] ( 31.94,149.08) --
	(213.30,149.08);

\path[draw=drawColor,line width= 0.5pt,line join=round] ( 81.41, 57.56) --
	( 81.41,191.52);

\path[draw=drawColor,line width= 0.5pt,line join=round] (163.84, 57.56) --
	(163.84,191.52);
\definecolor{fillColor}{RGB}{215,48,31}

\path[fill=fillColor] (176.20, 63.65) rectangle (200.93, 91.45);
\definecolor{fillColor}{RGB}{252,141,89}

\path[fill=fillColor] (151.47, 63.65) rectangle (176.20,119.14);
\definecolor{fillColor}{RGB}{253,204,138}

\path[fill=fillColor] (126.74, 63.65) rectangle (151.47,185.43);
\definecolor{fillColor}{gray}{0.60}

\path[fill=fillColor] ( 44.31, 63.65) rectangle (118.50, 98.83);
\definecolor{drawColor}{gray}{0.20}

\path[draw=drawColor,line width= 0.5pt,line join=round,line cap=round] ( 31.94, 57.56) rectangle (213.30,191.52);
\end{scope}
\begin{scope}
\path[clip] (  0.00,  0.00) rectangle (217.80,196.02);
\definecolor{drawColor}{gray}{0.30}

\node[text=drawColor,anchor=base east,inner sep=0pt, outer sep=0pt, scale=  0.72] at ( 27.89, 61.17) {0\%};

\node[text=drawColor,anchor=base east,inner sep=0pt, outer sep=0pt, scale=  0.72] at ( 27.89,103.89) {5\%};

\node[text=drawColor,anchor=base east,inner sep=0pt, outer sep=0pt, scale=  0.72] at ( 27.89,146.60) {10\%};
\end{scope}
\begin{scope}
\path[clip] (  0.00,  0.00) rectangle (217.80,196.02);
\definecolor{drawColor}{gray}{0.20}

\path[draw=drawColor,line width= 0.5pt,line join=round] ( 29.69, 63.65) --
	( 31.94, 63.65);

\path[draw=drawColor,line width= 0.5pt,line join=round] ( 29.69,106.37) --
	( 31.94,106.37);

\path[draw=drawColor,line width= 0.5pt,line join=round] ( 29.69,149.08) --
	( 31.94,149.08);
\end{scope}
\begin{scope}
\path[clip] (  0.00,  0.00) rectangle (217.80,196.02);
\definecolor{drawColor}{gray}{0.20}

\path[draw=drawColor,line width= 0.5pt,line join=round] ( 81.41, 55.31) --
	( 81.41, 57.56);

\path[draw=drawColor,line width= 0.5pt,line join=round] (163.84, 55.31) --
	(163.84, 57.56);
\end{scope}
\begin{scope}
\path[clip] (  0.00,  0.00) rectangle (217.80,196.02);
\definecolor{drawColor}{gray}{0.30}

\node[text=drawColor,anchor=base,inner sep=0pt, outer sep=0pt, scale=  0.72] at ( 81.41, 48.55) {Native};

\node[text=drawColor,anchor=base,inner sep=0pt, outer sep=0pt, scale=  0.72] at (163.84, 48.55) {Synthetic};
\end{scope}
\begin{scope}
\path[clip] (  0.00,  0.00) rectangle (217.80,196.02);
\definecolor{drawColor}{RGB}{0,0,0}

\node[text=drawColor,anchor=base,inner sep=0pt, outer sep=0pt, scale=  0.90] at (122.62, 38.70) {Icebergs};
\end{scope}
\begin{scope}
\path[clip] (  0.00,  0.00) rectangle (217.80,196.02);
\definecolor{drawColor}{RGB}{0,0,0}

\node[text=drawColor,rotate= 90.00,anchor=base,inner sep=0pt, outer sep=0pt, scale=  0.90] at ( 10.70,124.54) {(Iceberg / Total traded volume)};
\end{scope}
\begin{scope}
\path[clip] (  0.00,  0.00) rectangle (217.80,196.02);
\definecolor{fillColor}{RGB}{255,255,255}

\path[fill=fillColor] ( 42.01,  4.50) rectangle (203.23, 27.95);
\end{scope}
\begin{scope}
\path[clip] (  0.00,  0.00) rectangle (217.80,196.02);
\definecolor{drawColor}{RGB}{0,0,0}

\node[text=drawColor,anchor=base west,inner sep=0pt, outer sep=0pt, scale=  0.90] at ( 46.51, 13.13) {Min. no. tranches};
\end{scope}
\begin{scope}
\path[clip] (  0.00,  0.00) rectangle (217.80,196.02);
\definecolor{fillColor}{RGB}{255,255,255}

\path[fill=fillColor] (122.07,  9.00) rectangle (136.53, 23.45);
\end{scope}
\begin{scope}
\path[clip] (  0.00,  0.00) rectangle (217.80,196.02);
\definecolor{fillColor}{RGB}{253,204,138}

\path[fill=fillColor] (122.78,  9.71) rectangle (135.81, 22.74);
\end{scope}
\begin{scope}
\path[clip] (  0.00,  0.00) rectangle (217.80,196.02);
\definecolor{fillColor}{RGB}{255,255,255}

\path[fill=fillColor] (149.12,  9.00) rectangle (163.58, 23.45);
\end{scope}
\begin{scope}
\path[clip] (  0.00,  0.00) rectangle (217.80,196.02);
\definecolor{fillColor}{RGB}{252,141,89}

\path[fill=fillColor] (149.84,  9.71) rectangle (162.87, 22.74);
\end{scope}
\begin{scope}
\path[clip] (  0.00,  0.00) rectangle (217.80,196.02);
\definecolor{fillColor}{RGB}{255,255,255}

\path[fill=fillColor] (176.18,  9.00) rectangle (190.63, 23.45);
\end{scope}
\begin{scope}
\path[clip] (  0.00,  0.00) rectangle (217.80,196.02);
\definecolor{fillColor}{RGB}{215,48,31}

\path[fill=fillColor] (176.89,  9.71) rectangle (189.92, 22.74);
\end{scope}
\begin{scope}
\path[clip] (  0.00,  0.00) rectangle (217.80,196.02);
\definecolor{drawColor}{RGB}{0,0,0}

\node[text=drawColor,anchor=base west,inner sep=0pt, outer sep=0pt, scale=  0.72] at (141.03, 13.75) {3};
\end{scope}
\begin{scope}
\path[clip] (  0.00,  0.00) rectangle (217.80,196.02);
\definecolor{drawColor}{RGB}{0,0,0}

\node[text=drawColor,anchor=base west,inner sep=0pt, outer sep=0pt, scale=  0.72] at (168.08, 13.75) {4};
\end{scope}
\begin{scope}
\path[clip] (  0.00,  0.00) rectangle (217.80,196.02);
\definecolor{drawColor}{RGB}{0,0,0}

\node[text=drawColor,anchor=base west,inner sep=0pt, outer sep=0pt, scale=  0.72] at (195.13, 13.75) {5};
\end{scope}
\end{tikzpicture}}
  \hfill{}
  \subfigure[By number of orders.  Synthetic iceberg order tranches are counted either as one order, "s. (as one)", or separate orders, "s. (all tranches)"]{% Created by tikzDevice version 0.12 on 2019-08-28 16:06:03
% !TEX encoding = UTF-8 Unicode
\begin{tikzpicture}[x=1pt,y=1pt]
\definecolor{fillColor}{RGB}{255,255,255}
\path[use as bounding box,fill=fillColor,fill opacity=0.00] (0,0) rectangle (217.80,196.02);
\begin{scope}
\path[clip] (  0.00,  0.00) rectangle (217.80,196.02);
\definecolor{drawColor}{RGB}{255,255,255}
\definecolor{fillColor}{RGB}{255,255,255}

\path[draw=drawColor,line width= 0.5pt,line join=round,line cap=round,fill=fillColor] (  0.00,  0.00) rectangle (217.80,196.02);
\end{scope}
\begin{scope}
\path[clip] ( 28.35, 57.56) rectangle (213.30,191.52);
\definecolor{fillColor}{RGB}{255,255,255}

\path[fill=fillColor] ( 28.35, 57.56) rectangle (213.30,191.52);
\definecolor{drawColor}{gray}{0.92}

\path[draw=drawColor,line width= 0.2pt,line join=round] ( 28.35, 84.90) --
	(213.30, 84.90);

\path[draw=drawColor,line width= 0.2pt,line join=round] ( 28.35,127.40) --
	(213.30,127.40);

\path[draw=drawColor,line width= 0.2pt,line join=round] ( 28.35,169.90) --
	(213.30,169.90);

\path[draw=drawColor,line width= 0.5pt,line join=round] ( 28.35, 63.65) --
	(213.30, 63.65);

\path[draw=drawColor,line width= 0.5pt,line join=round] ( 28.35,106.15) --
	(213.30,106.15);

\path[draw=drawColor,line width= 0.5pt,line join=round] ( 28.35,148.65) --
	(213.30,148.65);

\path[draw=drawColor,line width= 0.5pt,line join=round] ( 28.35,191.15) --
	(213.30,191.15);

\path[draw=drawColor,line width= 0.5pt,line join=round] ( 63.02, 57.56) --
	( 63.02,191.52);

\path[draw=drawColor,line width= 0.5pt,line join=round] (120.82, 57.56) --
	(120.82,191.52);

\path[draw=drawColor,line width= 0.5pt,line join=round] (178.62, 57.56) --
	(178.62,191.52);
\definecolor{fillColor}{RGB}{215,48,31}

\path[fill=fillColor] (129.49, 63.65) rectangle (146.83,102.35);
\definecolor{fillColor}{RGB}{252,141,89}

\path[fill=fillColor] (112.15, 63.65) rectangle (129.49,130.42);
\definecolor{fillColor}{RGB}{253,204,138}

\path[fill=fillColor] ( 94.81, 63.65) rectangle (112.15,185.43);
\definecolor{fillColor}{RGB}{215,48,31}

\path[fill=fillColor] (187.29, 63.65) rectangle (204.63, 67.93);
\definecolor{fillColor}{RGB}{252,141,89}

\path[fill=fillColor] (169.95, 63.65) rectangle (187.29, 72.29);
\definecolor{fillColor}{RGB}{253,204,138}

\path[fill=fillColor] (152.61, 63.65) rectangle (169.95, 83.27);
\definecolor{fillColor}{gray}{0.60}

\path[fill=fillColor] ( 37.02, 63.65) rectangle ( 89.03, 65.08);
\definecolor{drawColor}{gray}{0.20}

\path[draw=drawColor,line width= 0.5pt,line join=round,line cap=round] ( 28.35, 57.56) rectangle (213.30,191.52);
\end{scope}
\begin{scope}
\path[clip] (  0.00,  0.00) rectangle (217.80,196.02);
\definecolor{drawColor}{gray}{0.30}

\node[text=drawColor,anchor=base east,inner sep=0pt, outer sep=0pt, scale=  0.72] at ( 24.30, 61.17) {0\%};

\node[text=drawColor,anchor=base east,inner sep=0pt, outer sep=0pt, scale=  0.72] at ( 24.30,103.67) {2\%};

\node[text=drawColor,anchor=base east,inner sep=0pt, outer sep=0pt, scale=  0.72] at ( 24.30,146.17) {4\%};

\node[text=drawColor,anchor=base east,inner sep=0pt, outer sep=0pt, scale=  0.72] at ( 24.30,188.67) {6\%};
\end{scope}
\begin{scope}
\path[clip] (  0.00,  0.00) rectangle (217.80,196.02);
\definecolor{drawColor}{gray}{0.20}

\path[draw=drawColor,line width= 0.5pt,line join=round] ( 26.10, 63.65) --
	( 28.35, 63.65);

\path[draw=drawColor,line width= 0.5pt,line join=round] ( 26.10,106.15) --
	( 28.35,106.15);

\path[draw=drawColor,line width= 0.5pt,line join=round] ( 26.10,148.65) --
	( 28.35,148.65);

\path[draw=drawColor,line width= 0.5pt,line join=round] ( 26.10,191.15) --
	( 28.35,191.15);
\end{scope}
\begin{scope}
\path[clip] (  0.00,  0.00) rectangle (217.80,196.02);
\definecolor{drawColor}{gray}{0.20}

\path[draw=drawColor,line width= 0.5pt,line join=round] ( 63.02, 55.31) --
	( 63.02, 57.56);

\path[draw=drawColor,line width= 0.5pt,line join=round] (120.82, 55.31) --
	(120.82, 57.56);

\path[draw=drawColor,line width= 0.5pt,line join=round] (178.62, 55.31) --
	(178.62, 57.56);
\end{scope}
\begin{scope}
\path[clip] (  0.00,  0.00) rectangle (217.80,196.02);
\definecolor{drawColor}{gray}{0.30}

\node[text=drawColor,anchor=base,inner sep=0pt, outer sep=0pt, scale=  0.72] at ( 63.02, 48.55) {Native};

\node[text=drawColor,anchor=base,inner sep=0pt, outer sep=0pt, scale=  0.72] at (120.82, 48.55) {S. (all tranches)};

\node[text=drawColor,anchor=base,inner sep=0pt, outer sep=0pt, scale=  0.72] at (178.62, 48.55) {S. (as one)};
\end{scope}
\begin{scope}
\path[clip] (  0.00,  0.00) rectangle (217.80,196.02);
\definecolor{drawColor}{RGB}{0,0,0}

\node[text=drawColor,anchor=base,inner sep=0pt, outer sep=0pt, scale=  0.90] at (120.82, 38.70) {Icebergs};
\end{scope}
\begin{scope}
\path[clip] (  0.00,  0.00) rectangle (217.80,196.02);
\definecolor{drawColor}{RGB}{0,0,0}

\node[text=drawColor,rotate= 90.00,anchor=base,inner sep=0pt, outer sep=0pt, scale=  0.90] at ( 10.70,124.54) {(\# Icebergs / Total \# orders)};
\end{scope}
\begin{scope}
\path[clip] (  0.00,  0.00) rectangle (217.80,196.02);
\definecolor{fillColor}{RGB}{255,255,255}

\path[fill=fillColor] ( 40.21,  4.50) rectangle (201.43, 27.95);
\end{scope}
\begin{scope}
\path[clip] (  0.00,  0.00) rectangle (217.80,196.02);
\definecolor{drawColor}{RGB}{0,0,0}

\node[text=drawColor,anchor=base west,inner sep=0pt, outer sep=0pt, scale=  0.90] at ( 44.71, 13.13) {Min. no. tranches};
\end{scope}
\begin{scope}
\path[clip] (  0.00,  0.00) rectangle (217.80,196.02);
\definecolor{fillColor}{RGB}{255,255,255}

\path[fill=fillColor] (120.27,  9.00) rectangle (134.73, 23.45);
\end{scope}
\begin{scope}
\path[clip] (  0.00,  0.00) rectangle (217.80,196.02);
\definecolor{fillColor}{RGB}{253,204,138}

\path[fill=fillColor] (120.98,  9.71) rectangle (134.01, 22.74);
\end{scope}
\begin{scope}
\path[clip] (  0.00,  0.00) rectangle (217.80,196.02);
\definecolor{fillColor}{RGB}{255,255,255}

\path[fill=fillColor] (147.32,  9.00) rectangle (161.78, 23.45);
\end{scope}
\begin{scope}
\path[clip] (  0.00,  0.00) rectangle (217.80,196.02);
\definecolor{fillColor}{RGB}{252,141,89}

\path[fill=fillColor] (148.04,  9.71) rectangle (161.07, 22.74);
\end{scope}
\begin{scope}
\path[clip] (  0.00,  0.00) rectangle (217.80,196.02);
\definecolor{fillColor}{RGB}{255,255,255}

\path[fill=fillColor] (174.38,  9.00) rectangle (188.83, 23.45);
\end{scope}
\begin{scope}
\path[clip] (  0.00,  0.00) rectangle (217.80,196.02);
\definecolor{fillColor}{RGB}{215,48,31}

\path[fill=fillColor] (175.09,  9.71) rectangle (188.12, 22.74);
\end{scope}
\begin{scope}
\path[clip] (  0.00,  0.00) rectangle (217.80,196.02);
\definecolor{drawColor}{RGB}{0,0,0}

\node[text=drawColor,anchor=base west,inner sep=0pt, outer sep=0pt, scale=  0.72] at (139.23, 13.75) {3};
\end{scope}
\begin{scope}
\path[clip] (  0.00,  0.00) rectangle (217.80,196.02);
\definecolor{drawColor}{RGB}{0,0,0}

\node[text=drawColor,anchor=base west,inner sep=0pt, outer sep=0pt, scale=  0.72] at (166.28, 13.75) {4};
\end{scope}
\begin{scope}
\path[clip] (  0.00,  0.00) rectangle (217.80,196.02);
\definecolor{drawColor}{RGB}{0,0,0}

\node[text=drawColor,anchor=base west,inner sep=0pt, outer sep=0pt, scale=  0.72] at (193.33, 13.75) {5};
\end{scope}
\end{tikzpicture}}
  \hfill{}
  \caption{Proportion of iceberg orders to all orders on one day.}
  \label{fig:Iceberg-proportions}
\end{figure}

\begin{figure}[H]
\centering{}%
\begin{minipage}[t]{0.5\textwidth}%
\begin{center}
\begin{center}
\input{fig/ice_20190618_peak_volume_dist.tex}
\par\end{center}\caption{\label{fig:Peak-volume-dist}Peak volume distribution.}
\par\end{center}%
\end{minipage}\hfill{}%
\begin{minipage}[t]{0.5\textwidth}%
\begin{center}
\begin{center}
\input{fig/ice_20190618_no_tranches_count.tex}
\par\end{center}\caption{\label{fig:Tranches-count}Number of tranches distribution.}
\par\end{center}%
\end{minipage}
\end{figure}

\begin{figure}[h]
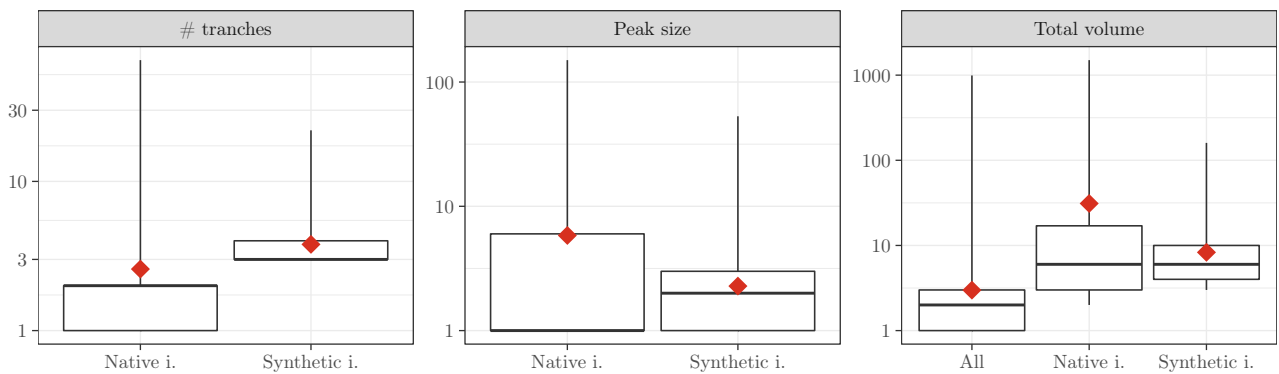

\centering{}\begin{center}
\include{fig/n_tranches_v_peak_volume_boxplot}
\par\end{center}\caption{\label{fig:volume-box-plot}Summary of the distributions of the number
of tranches per iceberg, the peak size and the total volume per order.
The lower and upper hinges correspond to the first and third quartiles.
The whiskers extend from the lower / upper hinge to the minimum /
maximum value, respectively. The middle bar is the median, while the
red diamond dot is the mean.}
\end{figure}

Figure \ref{fig:volume-box-plot} visualises summary statistics related
to the distributions of the number of tranches, the peak size and
the total volume per order. Note that the total volume of both native
and synthetic icebergs is significantly different from the the size
of all limit orders. Also, the median total volume is, in fact, identical
for native and synthetic icebergs (being equal to 6), but the means
are different due to some native icebergs having an extremely large
size.

\begin{figure}[H]
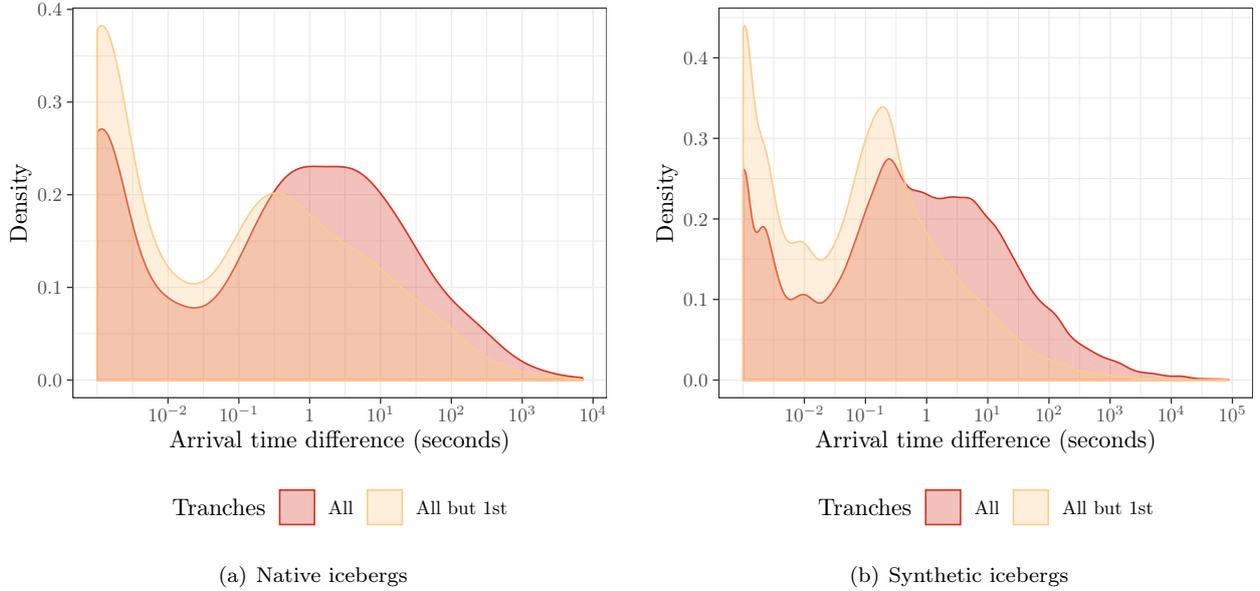

  \hfill{}
  \subfigure[Native icebergs]{\input{fig/ice_nat_20190618_atd_dist}}
  \hfill{}
  \subfigure[Synthetic icebergs]{\input{fig/ice_synth_20190618_atd_dist}}
  \hfill{}
  \caption{Tranche arrival time difference distributions (for values strictly greater than zero). It is instructive to compare two cases: when the initial tranche is included into and excluded from consideration~---~it might take longer time to execute the first tranche of an iceberg after its initial placement, but the following tranches get traded more rapidly.}
  \label{fig:atd-dist}
\end{figure}

Lastly, fig.~\ref{fig:atd-dist} shows the distribution of arrival
time differences between subsequent tranches. Zero values are discarded
for the purpose of drawing the plot, but they amount to 4.71\%\footnote{The fact that we observe zero delays for synthetic icebergs may be
attributed to an insufficient accuracy of time records (millisecond
resolution).} and 38.94\% of all values for synthetic and native icebergs, correspondingly.
If the initial tranche is not considered, then it can be seen that
the majority of tranches arrive less than one second after the previous
tranche (before being traded). This suggests that the proposed detection
algorithm is more suitable as an input to other trading algorithms,
rather than a signal to a day trader, who would not be able to react
sufficiently fast.

\subsection{Prediction Results}

After the model was trained, its predictive ability was evaluated
out of sample using the data for the period from 2019-06-19~16:45:00~CDT
to 2019-06-20~16:45:00~CDT. An extensive study of the model robustness
is required to claim its applicability on other instruments and time
spans; we leave this out of the scope of this article.

\paragraph{Synthetic icebergs}

The classification performance is summarised in tables~\ref{tab:Evaluation-metrics-synthetic}
and~\ref{tab:Confusion-matrices-synthetic}. $\hat{\mathcal{V}}^{\mathrm{all}}$
and $\hat{\mathcal{V}}^{\mathrm{unique}}$ volume aggregation methods
showed similar results (about 1\% difference in the classification
metrics). For the sake of brevity, only $\hat{\mathcal{V}}^{\mathrm{all}}$
and $\hat{\mathcal{V}}^{\mathrm{longest}}$ are displayed. The algorithm
demonstrates a fair performance as indicated by the~$\approx70\%$
accuracy, although it is mainly contributed by the true negatives
(the prediction that the iceberg is not complete). This is especially
true in the case of the ``all chains'' aggregation. When checking
for the equality $\hat{\mathcal{V}}_{r_{\max}}^{\mathrm{\cdot}}=\mathcal{V}_{r_{\max}}^{\cdot}$,
taking the longest chain gives a better result, because $\hat{\mathcal{V}}_{r_{\max}}^{\mathrm{longest}}$
is not averaged and thus is always an integer. It is instructive to
check the magnitude of the prediction error — the equality $\hat{\mathcal{V}}_{r_{\max}}^{\mathrm{\cdot}}=\mathcal{V}_{r_{\max}}^{\cdot}$
might not hold, but not by a large margin. Indeed, as the Regression
section of table~\ref{tab:Evaluation-metrics-synthetic} demonstrates,
the prediction is off by 1.78 units of volume on average.

\begin{table}[H]
\begin{centering}
\begin{tabular}{llll}
\toprule 
 &  & All chains average & Longest chain\tabularnewline
\midrule
\multirow{4}{*}{Classification} & Accuracy & 68.95\% & 67.54\%\tabularnewline
 & Precision & 52.66\% & 49.95\%\tabularnewline
 & Recall & 41.47\% & 83.67\%\tabularnewline
 & F1 score & 46.40\% & 62.56\%\tabularnewline
\midrule
\multirow{2}{*}{Regression} & MAE & 1.78 (22.55\%) & 2.15 (24.94\%)\tabularnewline
 & RMSE & 3.94 (49.85\%) & 4.52 (52.37\%)\tabularnewline
\bottomrule
\end{tabular}
\par\end{centering}
\caption{\label{tab:Evaluation-metrics-synthetic}Evaluation metrics for synthetic
icebergs. Percentages for regression are given relatively to the mean
total volume.}
\end{table}

\begin{table}[h]
\begin{centering}
{\small{}}%
\begin{tabular}{llllll}
\toprule 
 & \multicolumn{2}{c}{All chains average} &  & \multicolumn{2}{c}{Longest chain}\tabularnewline
\cmidrule{2-3} \cmidrule{3-3} \cmidrule{5-6} \cmidrule{6-6} 
 & Actual complete & Actual incomplete &  & Actual complete & Actual incomplete\tabularnewline
\midrule 
Predicted complete & 3621 (13.44\%) & 3255 (12.08\%) &  & 7305 (27.12\%) & 7318 (27.17\%)\tabularnewline
Predicted incomplete & 5110 (18.97\%) & 14952 (55.51\%) &  & 1426 (5.29\%) & 10889 (40.42\%)\tabularnewline
\bottomrule
\end{tabular}{\small\par}
\par\end{centering}
\caption{\label{tab:Confusion-matrices-synthetic}Confusion matrices for synthetic
icebergs.}
\end{table}

\paragraph{Native icebergs}

Of the total number of native icebergs (see fig. \ref{fig:Iceberg-completion-dist}),
33 icebergs with non-unique peak size values were filtered out, leaving
98\% of the initial amount used to estimate the total volume distribution.

Tables \ref{tab:Evaluation-metrics-native} and \ref{tab:Confusion-matrices-native}
summarise the predictive performance on native icebergs. Mode ($k$)
columns refer to the metrics and confusion matrices computed using
$k$ best mode predictions. For the sake of brevity, we only provide
median and mode (3) confusion matrices as those averages have demonstrated
the best results. As with synthetic icebergs, high accuracy values
are mainly contributed by a large number of true negatives. Note that
regression results are worse compared to the case of synthetic icebergs,
which can possibly be explained by the smaller sample size.

\begin{table}[H]
\begin{centering}
\begin{tabular}{lllllll}
\toprule 
 &  & {\footnotesize{}Mean} & {\footnotesize{}Median} & {\footnotesize{}Mode (1)} & {\footnotesize{}Mode (2)} & {\footnotesize{}Mode (3)}\tabularnewline
\midrule
\multirow{4}{*}{{\footnotesize{}Classification}} & {\footnotesize{}Accuracy} & {\footnotesize{}82.69\%} & {\footnotesize{}58.31\%} & {\footnotesize{}72.55\%} & {\footnotesize{}88.15\%} & {\footnotesize{}90.21\%}\tabularnewline
 & {\footnotesize{}Precision} & {\footnotesize{}33.33\%} & {\footnotesize{}19.22\%} & {\footnotesize{}26\%} & {\footnotesize{}73.03\%} & {\footnotesize{}83.91\%}\tabularnewline
 & {\footnotesize{}Recall} & {\footnotesize{}4.83\%} & {\footnotesize{}47.59\%} & {\footnotesize{}35.86\%} & {\footnotesize{}44.83\%} & {\footnotesize{}50.34\%}\tabularnewline
 & {\footnotesize{}F1 score} & {\footnotesize{}8.43\%} & {\footnotesize{}27.38\%} & {\footnotesize{}30.14\%} & {\footnotesize{}55.56\%} & {\footnotesize{}62.93\%}\tabularnewline
\midrule
\multirow{2}{*}{{\footnotesize{}Regression}} & {\footnotesize{}MAE} & {\footnotesize{}94.60 (97.87\%)} & {\footnotesize{}89.40 (92.49\%)} & {\footnotesize{}99.66 (103.14\%)} & {\footnotesize{}69.66 (72.07\%)} & {\footnotesize{}61.79 (63.92\%)}\tabularnewline
 & {\footnotesize{}RMSE} & {\footnotesize{}217.08 (224.58\%)} & {\footnotesize{}234.45 (242.55\%)} & {\footnotesize{}239.22 (247.49\%)} & {\footnotesize{}204.35 (211.41\%)} & {\footnotesize{}190.43 (197.01\%)}\tabularnewline
\bottomrule
\end{tabular}
\par\end{centering}
\caption{\label{tab:Evaluation-metrics-native}Evaluation metrics for native
icebergs. Percentages for regression are given relatively to the mean
total volume.}
\end{table}

\begin{table}[h]
\begin{centering}
{\small{}}%
\begin{tabular}{lllcll}
\toprule 
 & \multicolumn{2}{c}{Median} &  & \multicolumn{2}{c}{Mode (3)}\tabularnewline
\cmidrule{2-3} \cmidrule{3-3} \cmidrule{5-6} \cmidrule{6-6} 
 & Actual complete & Actual incomplete &  & Actual complete & Actual incomplete\tabularnewline
\midrule 
Predicted complete & 69 (7.85\%) & 290 (33.02\%) &  & 73 (8.31\%) & 14 (1.59\%)\tabularnewline
Predicted incomplete & 76 (8.65\%) & 443 (50.45\%) &  & 72 (8.2\%) & 719 (81.89\%)\tabularnewline
\bottomrule
\end{tabular}{\small\par}
\par\end{centering}
\caption{\label{tab:Confusion-matrices-native}Confusion matrices for native
icebergs.}
\end{table}

\section{Discussion and Future Work}

We have proposed an algorithm for detection and prediction of both
native and synthetic iceberg orders on CME. It can work with streaming
data or with pre-recorded data in a suitable format equally well.
The learning phase relies on a set of standard mathematical methods
which are simple enough, so that they can be invoked from a corresponding
statistical library or implemented from scratch. The detection results
agree with some of the ones found in the literature. The prediction
performance is fair given the simplicity of the model and can be improved
further.

\subsection{Detection}

Detection of native icebergs is straightforward as the information
disseminated by the exchange is sufficient to reliably determine the
sequence of tranches that constitute an iceberg order.

On the other hand, detecting synthetic icebergs is conceptually more
complicated and can only be attempted by relying on various heuristics.
One inherent limitation of the proposed model is that the next tranche
is expected to arrive earlier than any other limit orders for the
same price and volume combination after a trade. In network graph
terms, each node cannot have more than one child. This limitation
may possibly be overcome by considering more complex graphs where
each tranche is allowed to have more than one child, although at present
it is unclear how to proceed with consistent inference in that case.
The value of $dt$ parameter can be optimised using cross-validation.

That being said, for an end user interested in predictions, it may
not matter at all whether the detected limit orders are a part of
an iceberg order or not. Generally speaking, an \emph{order flow pattern}
gets detected, for which a satisfactory prediction can be made. This
information can in turn be used as an input to trading algorithms.

\subsection{Learning and prediction}

For learning, the employed Kaplan–Meier estimate is arguably too simplistic.
However, it improves the approach of \citep{Christensen2013}, since
it additionally accounts for the fact that many synthetic icebergs
get cancelled before being fully executed. As simple as it is, the
model does possess satisfactory predictive power and we hope that
the present results will serve as a baseline for predictions using
more advanced models.

There is much space for improvement of the learning and prediction
procedures. One possibility is to optimize algorithm parameters, e.g.
by using statistical techniques such as $k$-fold validation or bootstrap.
Alternatively, for synthetic icebergs a better choice would be to
utilize a semi-parametric relative risk (Cox) model and include covariates
into the analysis, which would make the prediction more accurate.
For native icebergs a variety of models are available as we are not
restricted to the methods of survival analysis.

\bibliographystyle{cbe}
\bibliography{../bibliography}

\end{document}